\documentclass[11pt]{article}
\usepackage{amsfonts,amsmath}
\usepackage{latexsym,epic,graphicx}
\usepackage{amssymb}

\makeatletter

\@addtoreset{equation}{section}
\makeatother

\newcommand{\be}{\begin{eqnarray}}
\newcommand{\ee}{\end{eqnarray}}
\newcommand{\beg}{\begin{eqnarray*}}
\newcommand{\eeg}{\end{eqnarray*}}

\newcommand{\nn}{\nonumber}
\newcommand{\Diff}{\textit{Diff}}
\newcommand{\Symp}{\textit{Symp}}
\newcommand{\Spin}{\textit{Spin}}
\newcommand{\vp}{\varphi}
\newcommand{\diag}{\text{diag}}

\newcommand{\e}{\varepsilon}
\newcommand{\ep}{\varepsilon}
\newcommand{\eps}{\epsilon}
\newcommand{\p}{\partial}
\newcommand{\G}{\Gamma}
\newcommand{\g}{\gamma}
\newcommand{\ca}{\tilde{\alpha}}
\newcommand{\cb}{\tilde{\beta}}
\newcommand{\cm}{\tilde{m}}
\newcommand{\cn}{\tilde{n}}
\newcommand{\cza}{\tilde{a}}
\newcommand{\czb}{\tilde{b}}
\newcommand{\czc}{\tilde{c}}
\newcommand{\czd}{\tilde{d}}
\newcommand{\cze}{\tilde{e}}

\newcommand{\cV}{\mathcal{V}}
\newcommand{\cP}{\mathcal{P}}
\newcommand{\cQ}{\mathcal{Q}}
\newcommand{\cM}{\mathcal{M}}
\newcommand{\cN}{\mathcal{N}}
\newcommand{\D}{\mathcal{D}} 

\newcommand{\dF}{\tilde{F}}
\newcommand{\dA}{\tilde{A}}

\newcommand{\Z}{\mathbb{Z}}
\newcommand{\R}{\mathbb{R}} 
\newcommand{\N}{\mathbb{N}} 
\newcommand{\C}{\mathbb{C}}

\newcommand{\id}{\mathbf{1\hspace{-2.9pt}l}}

\begin{document}

{\flushright IHES/P/09/02\\[9mm]}

\renewcommand{\thefootnote}{\fnsymbol{footnote}}

\begin{center}
{\LARGE \bf Generalized $E_{7(7)}$ coset dynamics and $D=11$ supergravity  }\\[1cm]
Christian Hillmann\footnote[3]{E-mail: \tt hillmann@ihes.fr} 
\\[5mm]
{\sl  Institut des Hautes \'Etudes Scientifiques\\
           35, Route de Chartres\\
      91440 Bures-sur-Yvette, France} \\[15mm]

\begin{tabular}{p{12cm}}
\hline\\\hspace{5mm}{\bf Abstract:}
The hidden on-shell $E_{7(7)}$ symmetry of maximal supergravity is usually discussed in a truncation from $D=11$ to four dimensions. In this article, we reverse the logic and start from a theory with manifest off-shell $E_{7(7)}$ symmetry inspired by West's coset construction. Following de Wit's and Nicolai's idea that a $4+56$ dimensional ``exceptional geometry'' underlies maximal supergravity, we construct the corresponding Lagrangian and the supersymmetry variations for the $56$ dimensional subsector. We prove that both the dynamics and the supersymmetry coincide with $D=11$ supergravity in a truncation to $d=7$ in the expected way.
\\\\\hline
\end{tabular}\\[9mm]
\end{center}

\renewcommand{\thefootnote}{\arabic{footnote}}
\setcounter{footnote}{0}

\begin{section}{Introduction}
Soon after the construction of the maximally supersymmetric $D=11$
gravity theory \cite{CJS78}, it was realized that this theory
exhibits an exceptional hidden symmetry $E_{7(7)}$ upon
dimensional reduction from $D=11$ to $d=4$
\cite{CJ79}. Much work has been devoted to reveal the origin of this hidden symmetry, which has e.g. led to the $E_{10(10)}$- and $E_{11(11)}$-conjectures \cite{DHN02,W01} for symmetries of M-theory. The dynamical origin of $E_{7(7)}$ has remained mysterious however. Another source of interest for $E_{7(7)}$ is its possible link to the improved ultraviolet properties of $\cN=8$ supergravity in $d=4$ \cite{AHCK08,BCDJK08,GRV06a,GRV06b,KK08}.\\

In this article, the conventional logic is reversed: Instead of starting with a diffeomorphism covariant theory in eleven dimensions exhibiting a hidden $E_{7(7)}$-symmetry upon a reduction to $d=4$, the starting point will be a manifestly $E_{7(7)}$-invariant theory in sixty dimensions with a hidden diffeomorphism symmetry in a reduction to $D=11$. This $4+56$ dimensional setting is already hinted at by the local $SO(3,1)\times SU(8)/\Z_2$ covariance of $D=11$ supergravity \cite{dWN86} and by BPS-extended supergravity \cite{dWN00}.\\

The dynamics will be fixed by group theoretical requirements. The generalized coset dynamics \`a la West \cite{W00} will provide a general multi-parameter class of Lagrangians with manifest $E_{7(7)}$-invariance in $4+56$ dimensions. Then, we will uniquely fix a specific Lagrangian within this $E_{7(7)}$-invariant class by requiring the symmetry group $Gl(7)\subset E_{7(7)}$ to be enlarged to $\Diff(7)$ in a truncation to $4+7$ dimensions. The resulting theory will be shown to completely agree with the truncation of the bosonic part of $D=11$ supergravity to seven dimensions, if Cremmer \& Julia's identification of the $E_{7(7)}/(SU(8)/\Z_2)$ coset \cite{CJ79} with fields of supergravity is used and if we restrict supergravity to these degrees of freedom. \\

A similar analysis can be performed for fermions. In a first step, the introduction of an $SU(8)$ covariant derivation $\underline{\delta}$ with a $32$ dimensional Gra\ss mann valued parameter $\eps$ on the bosonic degrees of freedom in $4+56$ dimensions leads to Majorana fermions $\chi$. As for the bosonic Lagrangian, the general $SU(8)$ covariant derivation $\underline{\delta}$ of the fermions $\chi$ is not unique a priori. Requiring $\Diff(7)$ to appear as a hidden symmetry completely fixes this arbitrariness however. Using Cremmer \& Julia's redefinition of the gravitino $\psi$ \cite{CJ79}, these transformations are then found to agree with the supersymmetry transformations of $D=11$ supergravity subject to the same restrictions as used for the bosonic sector. Furthermore, the fermionic dynamics of $D=11$ supergravity can also be reproduced in the same way.\\

The geometrical setting for this generalized coset dynamics consists of a vector bundle $E$ with a $56$ dimensional fibre $E_x$ over a four-dimensional manifold with structure group $E_{7(7)}$. Following de Wit \& Nicolai \cite{dWN00}, it will be referred to as ``exceptional geometry''. Hence, it is obvious that the comparison with $D=11$ supergravity in the truncation to seven dimensions is the complicated part. The way how to compare the dynamics in the four base dimensions of exceptional geometry to supergravity is uniquely determined. It will be discussed elsewhere. The results of the present article hence provide only a partial proof of the statement that all solutions of $D=11$ supergravity form the subset of solutions of sixty dimensional exceptional geometry with $49$ independent Killing vectors.\newpage

The motivation for this construction is the following: The supersymmetry of $60$ dimensional exceptional geometry would immediately provide a Lagrangian of $\cN=8$ $d=4$ supergravity with manifest off-shell $E_{7(7)}$-invariance by dimensional reduction. It would also be interesting to further investigate the relation to Hull's and Waldram's constructions with vector bundles of $E_{7(7)}$-structure \cite{Hu07,PW08}.\\

The article is organized as follows: We will start with a brief introduction to the hidden symmetries of maximal supergravity in section \ref{SUGRA} that serve as a motivation for the $4+56$ dimensional setting as explained in section \ref{KK}. In section \ref{CosDyn}, the generalized coset dynamics is introduced and a symmetry enlargement is illustrated by the familiar example of the Einstein--Hilbert action. We will focus on the $E_{7(7)}/(SU(8)/\Z_2)$ coset in section \ref{BosDyn} and compare the bosonic Lagrangian with supergravity before discussing supersymmetry transformations and the fermionic dynamics in section \ref{SUSY}. Finally, the generalized coset dynamics will be related to the sixty dimensional exceptional geometry in section \ref{Complete}.
\end{section}

\begin{section}{Supergravity in eleven dimensions}
\label{SUGRA}
\begin{subsection}{Conventions}
\label{DynSUSY}
Following Nahm's result \cite{N77}, $\cN=8$ supergravity is referred to as maximal supergravity in $d=4$. This theory was first constructed from a Kaluza--Klein reduction of Cremmer, Julia \& Scherk's supergravity in $D=11$ \cite{CJS78}, which is provided by the following action with $\ca_i=0,\dots,10$
\be\label{Actionneu}
	S &=& \int\limits_{\cM^{11}}d^{11}x\det(E)\left(\frac{1}{4}\tilde{R}_{11} -\frac{1}{2}\bar{\psi}_{\ca_1}\tilde{\G}^{\ca_1\dots \ca_3}\nabla_{\ca_2}\psi_{\ca_3} -\frac{1}{48}F_{\ca_1\dots \ca_4}F^{\ca_1\dots \ca_4}\nn
	\right.\\
	&& -\frac{1}{96}\left( \bar{\psi}_{\ca_5}\tilde{\G}^{\ca_1\dots \ca_6}\psi_{\ca_6} +12\bar{\psi}^{\ca_1}\tilde{\G}^{\ca_2\ca_3}\psi^{\ca_4}\right)F_{\ca_1\dots \ca_4} \nn\\
	&&\left. +\frac{2}{12^4}\e^{\ca_1\dots \ca_{11}}F_{\ca_1\dots \ca_4}F_{\ca_5\dots \ca_8}A_{\ca_9\dots \ca_{11}}\right).
\ee
To simplify the notation, we have absorbed the gravitational constant $\kappa$ in the fields as suggested in \cite{CJ79} and neglected both quartic terms in fermions and the quadratic ones that arise from the non-vanishing torsion of supergravity throughout the paper.\\

The metric degrees of freedom are encoded in the \textit{rep\`ere mobile} or \textit{vielbein} ${E_{\tilde{\mu}}}^{\ca}$ by the standard identification
\be\label{VielbeinG}
g&=& g_{\tilde{\mu}\tilde{\nu}}\,dx^{\tilde{\mu}} \otimes dx^{\tilde{\nu}}\nn\\
g_{{\tilde{\mu}}{\tilde{\nu}}} &=:&{E_{\tilde{\mu}}}^{\ca}{E_{\tilde{\nu}}}^{\cb}\eta_{\ca\cb}
\ee
with the signature $(-,+,\dots,+)$ of $\eta$. This is always possible given a manifold $\cM^{11}$ having vanishing first and second Stiefel--Whitney class and the metric $g$ being non-degenerate, what we henceforth assume. The definition of the vielbein (\ref{VielbeinG}) introduces an additional symmetry: a local Lorentz symmetry $O\in SO(10,1)$\footnote{Groups will be denoted by capital letters and their associated algebras by gothic ones.
}
\be\label{lokalLo}
{E'_{\tilde{\mu}}}^{\ca} &=&{E_{\tilde{\mu}}}^{\cb}{O_{\cb}}^{\ca}.
\ee
The local Lorentz (``flat'') indices $\ca_i=0,\dots,10$ in the action $S$ (\ref{Actionneu}) are hence raised and lowered with the flat Minkowski metric $\eta$, because we have transformed the three-form potential $A$, its corresponding field strength
\be\label{F4a}
F_{\ca_1\dots \ca_4} &:=&4\nabla_{[\ca_1}A_{\ca_2\dots \ca_4]}
\ee
and the gravitino $\psi$ into the vielbein frame by contraction with the inverse vielbein ${E_{\ca}}^{\mu}$.\footnote{Following the standard convention, we denote the inverse vielbein by a simple change of the indices $\delta_{\ca}^{\cb} = {E_{\tilde{\mu}}}^{\cb} {E_{\ca}}^{\tilde{\mu}}$ and introduce the abbreviation $\p_{\ca}:={E_{\ca}}^\mu\frac{\p}{\p x^\mu}$.
} 
Since torsion terms have been neglected, the covariant derivative $\nabla$ in the action (\ref{Actionneu}) is the standard Levi--Civita connection in the vielbein frame, sometimes referred to as spin connection.\footnote{We emphasize that the ``covariant derivative'' $D_{[\ca_2}\psi_{\ca_3]}$ in \cite{CJS78,FN76} only differs from $\nabla_{[\ca_2}\psi_{\ca_3]}$ by the torsion tensor that is not discussed in this article.} The Ricci scalar $\tilde{R}_{11}$ is composed of the vielbein $E$ by the following formul\ae:
\begin{subequations}\label{Abk}
\be
	\tilde{R}_{11} &:=& \eta^{\ca\cb}
	\left(2\p_{[\ca}  {\omega_{\tilde{\gamma}]\cb}}^{\tilde{\gamma}} 
+2 {\omega_{[\ca {\tilde{\gamma}}]}}^{\tilde{\delta}}{\omega_{{\tilde{\delta}}\cb}}^{\tilde{\gamma}}
+2{\omega_{[\ca|\cb|}}^{\tilde{\delta}} {\omega_{{\tilde{\gamma}}]{\tilde{\delta}}}}^{\tilde{\gamma}}\right)
\\
\omega_{\ca \cb \tilde{\gamma}} &:=& (Q_{\ca})_{\cb \tilde{\gamma}} -2(P_{[\cb})_{\tilde{\gamma}]\ca}
\\
(Q_{\ca})_{\cb \tilde{\gamma}}&:=& \eta_{\tilde{\delta}[\tilde{\gamma}}{E_{\cb]}}^\mu\p_{\ca} {E_\mu}^{\tilde{\delta}}
\\
(P_{\ca})_{\cb \tilde{\gamma}}&:=& \eta_{\tilde{\delta}(\tilde{\gamma}}{E_{\cb)}}^\mu\p_{\ca} {E_\mu}^{\tilde{\delta}}.
\ee
\end{subequations}

Furthermore, we use the real matrix representation $\tilde{\G}^{\ca}\in \R^{32\times 32}$ of the Clifford algebra $\{\tilde{\G}^{\ca},\tilde{\G}^{\cb}\}=2\eta^{\ca\cb}$ with normalization $\tilde{\G}^{\ca_1\dots \ca_{11}}=\e^{\ca_1\dots \ca_{11}}\id_{32}$ and $\e^{0\,1\,2\,3\,4\,5\,6\,7\,8\,9\,10}=1$. We have suppressed the spinor indices of $\psi$ and $\tilde{\G}$ and introduced the standard abbreviation for the real Majorana conjugate spinor
\beg
\bar{\psi}^{\ca}&:=&\big(\psi^{\ca}\big)^t\G^0.
\eeg

In this paper, we will discuss theories on the level of the Lagrangian. The only equation of motion that will be necessary is the one of the four-form field strength $F$, which, to zeroth order in fermions, reads
\beg
\nabla_{\ca_0}F^{\ca_0\dots \ca_3} &=& -\frac{1}{24^2}\e^{\ca_1\dots \ca_{11}} F_{\ca_4\dots \ca_7}F_{\ca_{8}\dots \ca_{11}}.
\eeg
In a first order formalism, it is equivalent \cite{W00} to the two equations
\begin{subequations}\label{4FormG}
\be\label{4Form}
F^{\ca_1\dots \ca_4} &=:& \frac{1}{7!} \e^{\ca_1\dots \ca_{11}}\dF_{\ca_4\dots \ca_{11}}\\
\dF_{\ca_1\dots \ca_7} &=& 7\left(\nabla_{[\ca_1}\dA_{\ca_2\dots \ca_7]} + 5A_{[\ca_1\dots \ca_3}F_{\ca_4\dots \ca_7]}\right),
\label{7Form}
				\ee
		\end{subequations}
with a dual six-form potential $\dA$ and its corresponding seven-form field strength $\dF$. The set of solutions of the equations of motion is invariant under the following supersymmetry transformations that are modulo non-linear terms in fermions:
\begin{subequations}\label{Trafo1neu}
\be
	{E_{\ca}}^{\tilde{\mu}} \delta_{\ep} {E_{\tilde{\mu}}}^{\cb} &=& \bar{\ep}\tilde{\G}^{\cb}\psi_{\ca}
	\\
	\delta_{\ep} \psi_{\cb} &=& \nabla_{\cb}\ep +\frac{1}{144}\left(\tilde{\G}{{}^{\ca_1\dots \ca_4}}_{\cb}-8\delta_{\cb}^{\ca_1}\tilde{\G}^{\ca_2\dots \ca_4}\right)\ep F_{\ca_1\dots \ca_4} \\
	\delta_{\ep} A_{\ca_1\dots \ca_3} &=& -\frac{3}{2}\bar{\ep}\tilde{\G}_{[\ca_1\ca_2}\psi_{\ca_3]}.
\ee
\end{subequations}
All fields $(E,A,\psi,\ep)$ are manifestly real in the present conventions. We want to close this section with the well-known fact \cite{CJS78} that the supersymmetry algebra only closes on-shell. This is encoded in the following equivalence relation modulo the equations of motion:
\beg
\left[\delta_{\ep_1},\delta_{\ep_2}\right] &\sim& \delta_{\ep_3} + \delta_{\mathfrak{diff}_{11}} + \delta_{\mathfrak{so}_{(10,1)}} + \delta_{3\text{-form gauge}}.
\eeg
Neither an off-shell formulation of the supersymmetry algebra acting on the fields $(E,A,\psi)$ nor an unconstrained superspace formulation of $D=11$ supergravity has been constructed so far.\footnote{Constraints are an essential ingredient in Cremmer \& Ferrara's construction \cite{CF80}. Another interesting formulation is the light-cone approach \cite{BKR08}.} In particular, a combination of the $\mathfrak{so}_{(10,1)}$ representations $(E,A,\psi)$ into independent representations of some superalgebra that is a symmetry of the equations of motion has not been achieved yet. We will not use these concepts in this article, but rather focus on the hidden symmetries. We shall see that $E_{7(7)}$ suggests a complementary way to discuss the independent degrees of freedom of supergravity.
\end{subsection}

\begin{subsection}{Hidden symmetries}
\label{Hid}
A Kaluza--Klein reduction\footnote{For this article, a ``Kaluza--Klein reduction'' implies a truncation of the massive modes.} of $D=11$ supergravity on a flat spacelike hypertorus $T^n$ for $n=1,\dots,9$ is equivalent to restricting the set of solutions to the ones with $n$ independent, spacelike, commuting Killing vectors. These sets of solutions are orbits of the symmetry groups $E_{n(n)}$ \cite{J81,J83,N87} that are called ``hidden symmetries'', because their origin is not obvious from the action of $D=11$ supergravity in the form stated in (\ref{Actionneu}). It is remarkable that even for the reduced supergravity to four space time dimensions, it has not been possible to construct an action with manifest $E_{7(7)}$-invariance so far.
\\

In this article, we will focus on the role of the $133$ dimensional symmetry group $E_{7(7)}$  with its maximal compact subgroup $SU(8)/\Z_2$ \cite{CJ79} in the {\textit{unreduced}} $D=11$ supergravity. From the dynamical point of view, a very interesting result addressing this question was established by de Wit \& Nicolai in 1986:
\begin{center}
 \textit{$SO(3,1)\times SU(8)/\Z_2$ is a local symmetry of the equations of motion of $D=11$ supergravity \cite{dWN86}.}
\end{center}
Guided by the discovery of the global $E_{7(7)}$ symmetry in $d=4$ $\cN=8$ supergravity \cite{CJ79}, their ansatz reduced the manifest local Lorentz symmetry $SO(10,1)$ (\ref{lokalLo}) to $SO(3,1)\times SO(7)$ by fixing a particular matrix form for the vielbein:
	\be\label{Vielbdecom}
	{E_{\tilde{\mu}}}^{\ca}
	&=:&
	{\left(\begin{tabular}{c|c}
	$\Delta^{-\frac{1}{2}}e_{\mu}{}^{\alpha}$ & $B_{\mu}{}^a$\\
	\hline
	$0$												& ${e_m}^a$
	\end{tabular}
	\right)_{\tilde{\mu}}}^{\ca}\\
\text{with}\quad 
\Delta &:=& \det\left({e_m}^a\right)
\label{sigmaDefi2}
\\
\text{and}\quad \mu,\alpha &=& 0,\dots,3,\nn\\
m,a &=& 4,\dots,10,\nn\\
\tilde{\mu},\ca &=& 0,\dots,10.\nn
	\ee
	Then they combined the degrees of freedom of the vielbein and of the three-form potential into $SO(3,1)\times SU(8)/\Z_2$ representations in such a way that both the supersymmetry variations and the equations of motion of $D=11$ supergravity exhibit manifest local $\Spin(3,1)\times SU(8)$ covariance.\footnote{Due to the presence of the fermions it is necessary to pass to the covering group as usual. De Wit \& Nicolai used the same redefinition for the fermions as Cremmer \& Julia in \cite{CJ79} that we will also use in section \ref{SUSY}.
}
\end{subsection}
\end{section}

\begin{section}{Motivation from Kaluza--Klein theory}
\label{KK}
In this article, we would like to present a different interpretation of this $SO(3,1)\times SU(8)/\Z_2$ symmetry of $D=11$ supergravity. This is related to the following observation in Kaluza--Klein theory:\\

In $d+1$ dimensional pure gravity, the metric or equivalently, the vielbein $E$ (\ref{VielbeinG}) is the only independent field. The action is provided by the standard Einstein--Hilbert action
\be\label{EH}
	S_{EH} &=& \int\limits_{\cM^{d+1}}d^{d+1}x\det(E)\,\frac{1}{4}\tilde{R}_{d+1}
\ee
with the obvious generalization of (\ref{Abk}) to the Ricci scalar $\tilde{R}_{d+1}$ in $d+1$ dimensions. Substituting these explicit expressions in terms of the vielbein $E$ produces a Lagrangian that only contains the vielbein, its inverse, the Minkowski metric $\eta$ and partial derivatives $\frac{\p}{\p x^{\cm}}$ with $\cm=0,\dots,d$ in the coordinate induced frame. In this notation, the symmetry of the theory under the following two transformations is obvious:
\begin{enumerate}
	\item General coordinate transformations $\vp\in \Diff(d+1)$ and
	\item local Lorentz transformations $O\in SO(d,1)$
\end{enumerate}
that act as follows:
	\be\label{vielb6}
	\frac{\p\vp^{\cm}}{\p x^{\cn}}{E_{\cm}'}^{\cza} &=& {E_{\cn}}^{\czb} \, {O_{\czb}}^{\cza}.
	\ee
	The indices $\cm,\cn,\cza,\czb$ take values in $0,\dots,d$. A reduction \`a la Kaluza--Klein $d+1\rightarrow d$ amounts to choosing one coordinate on which the field does not depend. Since the vielbein ${E_\mu}^a$ is the only dynamical field in $d+1$ pure gravity, we can without loss of generality impose that it does not depend on the $d^{\text{th}}$ coordinate, i.e.
\be\label{reduction}
\frac{\p}{\p x^d} \left({E_\mu}^a\right)=0.
\ee
Since all coordinate indices must be contracted with partial derivatives $\frac{\p}{\p x^{\cm}}$, the constraint (\ref{reduction}) reduces the effective range of the indices in the Lagrangian to
\begin{enumerate}
	\item coordinate indices:\quad $\cm\,=\,0,\dots,d-1$ \quad and
	\item vielbein indices:\quad \quad\,$\cza\,=\,0,\dots,d$.
\end{enumerate}
The symmetries that keep the reduction (\ref{reduction}) invariant are
\begin{enumerate}
	\item General coordinate transformations $ \Diff(d)\times Gl(1)$ and
	\item local Lorentz transformations $SO(d,1)$.
\end{enumerate}
The important fact is that the local Lorentz transformations $SO(d,1)$ are not affected by a Kaluza--Klein reduction a priori. It is possible and conventional to also reduce $SO(d,1)$ to $SO(d-1,1)$ by fixing a particular matrix form of the vielbein in $d+1$ dimensions similarly to the reduction of $SO(10,1)$ to $SO(3,1)\times SO(7)$ in (\ref{Vielbdecom}), but this is not compulsory. Hence, it is perfectly consistent to discuss a $d$ dimensional theory with general coordinate symmetry $\Diff(d)$ and enlarged local Lorentz symmetry $SO(d,1)$.\\

To establish the connection to $D=11$ supergravity, we should recall that de Wit \& Nicolai in fact enlarged the remaining local Lorentz symmetry $SO(3,1)\times SO(7)$ to $SO(3,1)\times SU(8)/\Z_2$ in \cite{dWN86}. Hence, $D=11$ supergravity can be viewed as an eleven dimensional theory with enlarged local Lorentz-like symmetry $SO(3,1)\times SU(8)/\Z_2$. In this article, we shall investigate the consequences of interpreting $SO(3,1)\times SU(8)/\Z_2$ as the local Lorentz symmetry of a higher dimensional space that leads to supergravity in a reduction to $D=11$ dimensions.\\

Thus, we are led to the question of finding the lowest dimensional Lorentz group $SO(d-1,1)$ with the property
\beg
SO(3,1)\times SU(8)/\Z_2 &\subset & SO(d-1,1).
\eeg
The answer is provided by representation theory. Due to the division by $\Z_2$, only $\mathfrak{su}_8$-representation vector spaces with an even number of $\mathfrak{su}_8$ indices also are representations of $SU(8)/\Z_2$. Furthermore, these are complex vector spaces, which leads to an additional factor $2$ for the real dimension:
\beg
d &\geq& 4+2\cdot \binom{8}{2} \,=\,60.
\eeg
This indicates that a $60$-dimensional structure may be relevant for maximal supergravity. However, it is well-known that there are two severe problems with discussing a sixty dimensional supergravity theory in the conventional setting:
\begin{enumerate}
	\item The supersymmetry parameter $\ep$ would have to transform as a representation of $\Spin(59,1)$ which would lead to more than $32$ supercharges in a compactification to $d=4$.
	\item The minimal number of off-shell degrees of freedom of a conventional gravitational theory in $d=60$ would be $\frac1260(60+1)$, much more than in $D=11$ supergravity.
\end{enumerate}
To sum up, there is little hope to match the dynamics of $D=11$ supergravity in a Kaluza--Klein reduction of an arbitrary sixty dimensional geometry. However, there is a way around these problems if an ``exceptional geometry'' is adapted. Using the tool of generalized coset dynamics, we will make this more precise in the following sections.
\end{section}

\begin{section}{Coset dynamics}
\label{CosDyn}
We will start by reviewing the conventional coset dynamics, before proposing its extension that will prove to be relevant for supergravity.

\begin{subsection}{Conventional coset dynamics}
\label{ConvCos}
The dynamical degrees of freedom in these theories are parametrized by a symmetric space, which is without loss of generality a right group coset
\be\label{Coset}
\cV &\in&  G/K
\ee
for a real, finite dimensional Lie group $G$ and its maximal compact subgroup $K$. A priori, there are independent left and right actions by $g\in G$ and $k\in K$ respectively:
\be\label{CosTrafo}
\cV\,' &=& g\cdot \cV \cdot k.
\ee
Note that this transformation (\ref{CosTrafo}) shows great similarity to the law of transformation of a vielbein under a combined $\Diff\,\times SO$ action (\ref{vielb6}). Passing to a matrix representation $\mathbf{R}$ of $G$ with $\mathbf{R}(g)\in \R^{d\times d}$, equation (\ref{CosTrafo}) reads with $\cm,\cn,\cza,\czb=1,\dots,d$
\beg
\mathbf{R}(\cV')_{\cm}{}^{\cza} &=& \mathbf{R}(g)_{\cm}{}^{\cn} \,\mathbf{R}(\cV)_{\cn}{}^{\czb}\, \mathbf{R}(k)_{\czb}{}^{\cza}.
\eeg
The coset element $\cV$ (\ref{Coset}) then corresponds to the vielbein by $\mathbf{R}(\cV)_{\cm}{}^{\cza}=E_{\cm}{}^{\cza}$, the left action to the Jacobi matrix by $\mathbf{R}(g^{-1})_{\cn}{}^{\cm}=\frac{\p \vp^{\cm}}{\p x^{\cn}}$ and the right action to the local Lorentz rotation $\mathbf{R}(k)_{\czb}{}^{\cza}= O_{\czb}{}^{\cza}$ in (\ref{vielb6}). This correspondence will be essential for the generalized coset dynamics of section \ref{GenCos}.\\

In complete analogy to fixing the local Lorentz symmetry by decreeing a particular matrix form for the vielbein (\ref{Vielbdecom}), it is possible to link the right $K$-action to the left $G$-action on $\cV$ by fixing the presentation of $\cV$, e.g. by a triangular gauge choice that is equivalent to decreeing that the matrix representation of $\cV$ be of triangular shape. Since a left $G$-action perturbs this setting in general, a compensating $k_g(\cV)\in K$ is needed to restore the shape of the matrix:
\be\label{CosTrafo2}
\cV\,' &=& g\cdot \cV \cdot k_g(\cV).
\ee
This is referred to as a non-linear realization of the symmetry $G$ on the symmetric space parametrized by the $\dim(g)-\dim(K)$ degrees of freedom of $\cV\in G/K$. It is important to keep in mind that the uniquely determined compensating rotation $k_g$ depends on $\cV$ in general.
\\

As a next step, assume that $\cV$ depends on some coordinates $x^{\cm}$ with $\cm=1,\dots,d$. Denoting the corresponding Lie algebr\ae{} by $\mathfrak{g}$ and $\mathfrak{k}$, the Maurer--Cartan form allows for the decomposition
\be\label{MCf}
\cV^{-1}\cdot d\cV &=& \cP + \cQ
\ee
with one-form valued objects 
\beg
\cP&=:& \cP_{\cm} dx^{\cm}\,\in\,\mathfrak{g}\ominus \mathfrak{k}\\
\cQ&=:& \cQ_{\cm} dx^{\cm}\,\in\, \mathfrak{k}.
\eeg
The transformation of the gauged fixed coset element $\cV$ under a left global action $g\in G$ (\ref{CosTrafo2}) dictates the induced transformation of $\cP$ and $\cQ$:
\begin{subequations}\label{PQTrafo}
\be
\cP' &=& k_g^{-1}\cdot \cP\cdot k_g\label{PQTrafo1}\\
\cQ' &=& k_g^{-1}\cdot \cQ\cdot k_g + k_g^{-1}\cdot d k_g.
\ee
\end{subequations}
Due to its transformation, $\cQ$ defines a covariant derivative $\nabla$ acting on $K$-representation spaces $\psi$ in a representation $\mathbf{R}$:
\be\label{covder}
\D\psi &:=&d\psi -\mathbf{R}(\cQ)\psi.
\ee
The one form $\cP$ however, transforms as a tensor (\ref{PQTrafo1}). It is therefore the basic building block of Lagrangians, such as
\be\label{Lagr1}
\mathcal{L} &=& g^{\mu\nu}\left\langle \cP_{\mu},\cP_{\nu}\right\rangle.
\ee
Here, $g^{\mu\nu}$ is the inverse of the relevant space-time metric $g$ (\ref{VielbeinG}) and $\langle\cdot,\cdot\rangle$ the Cartan--Killing metric of the Lie algebra $\mathfrak{g}$ that is proportional to the trace for the matrix representation of $\mathfrak{g}$. For $G=E_{7(7)}$ and four dimensional space-time $\mu,\nu=0,\dots,3$, the Lagrangian (\ref{Lagr1}) describes the dynamics of the scalar sector of $\cN=8$ $d=4$ supergravity \cite{CJ79}.\\

It is obvious that the set of possible Lagrangians that are quadratic in derivatives is quite restricted. For Lie groups with traceless matrix representations such as $G=E_{7(7)}$, the Lagrangian (\ref{Lagr1}) in fact is unique. The generalized coset dynamics of the next section will provide a wider choice of Lagrangians with $G=E_{7(7)}$-invariance.
\end{subsection}

\begin{subsection}{Generalized coset dynamics}
\label{GenCos}
Inspired by the pioneering work of Borisov \& Ogievetsky \cite{BO74} and West \cite{W00}, we would like to discuss the following extension. Assume that the coordinates $x^{\cm}$ the coset element $\cV$ depends on, form a {\sl representation space of $G$ of dimension $d$}. Introducing a matrix representation $\mathbf{R}(g)\in \R^{d\times d}$ for $g\in G$, one can without loss of generality define an action of $g\in G$ on the coordinates $x^{\cm}$ with $\cm,\cn =1,\dots,d$
\be\label{xTrafo}
x'{}^{\cm} &=&  \mathbf{R}(g^{-1})_{\cn}{}^{\cm} x^{\cn}.
\ee
As in section \ref{ConvCos}, we insist that $g\in G$ does not depend on the coordinates $x^{\cm}$, i.e. it is a global symmetry. Then the partial derivatives transform in the dual representation:
\be\label{xTrafo2}
\left(\frac{\p}{\p x^{\cn}}\right)' &=& {\mathbf{R}(g)_{\cn}}^{\cm}\frac{\p}{\p x^{\cm}}.
\ee
Hence, it is possible to define a derivative by multiplying with $\cV^{-1}$ that transforms by the induced $k_g\in K$ action under a global $G$ action (\ref{CosTrafo2}):
\be\label{underl}
\underline{\p}_{\cza}&:=& \mathbf{R}(\cV^{-1})_{\cza}{}^{\cm}\frac{\p}{\p x^{\cm}}\\
\text{with}\quad\underline{\p}_{\cza}' &=& \mathbf{R}(k_g^{-1})_{\cza}{}^{\czb}\, \underline{\p}_{\czb}
\label{pTrafo}
.
\ee 
The names of the indices of the matrix representations $\mathbf{R}$ are completely arbitrary a priori. It is only to emphasize the different transformation behaviour that we will follow the convention to use indices from the middle of the alphabet for objects that transform as $G$-representations (\ref{xTrafo2}) and indices from its beginning for $K\subset G$-representations such as (\ref{pTrafo}).\\

The definition of the derivative $\underline{\p}$ is the main ingredient of generalized coset dynamics: A short look at the transformation of $\cP$ (\ref{PQTrafo}) shows that it is consistent in this setting to contract the indices of $\underline{\p}$ with the coset indices of $\cP$ in a $K$-covariant way in order to construct a Lagrangian with $G$-invariance.\\

This construction seems to suffer from one drawback. If derivative indices are contracted with coset indices, the symmetry of general coordinate transformations $\Diff(d)$ is broken in general to the subgroup $G\subset Gl(d)$, because global transformations of the coordinates correspond to the $Gl(d)$ subgroup of $\Diff(d)$ \cite{Hi08}.\footnote{To be precise, it is broken to the affine group $A(d)$ being the semidirect product of $Gl(d)$ with the abelian group of translations whose Jacobi matrix is trivial, however. Hence, their induced $k_g\in K$ action is trivial, too.
} Keeping in mind the discussion at the end of section \ref{KK}, this exactly is what we want: An unbroken $\Diff(d)$-symmetry for $d=60$ would be inconsistent with maximal supergravity.\\

However, we want to establish contact with $D=11$ supergravity in the end, whose symmetry group contains $\Diff(11)$. Therefore, the constraint on the coupling of the Lagrangians must be that in a Kaluza--Klein reduction to eleven dimensions, the diffeomorphism covariance must be restored. Before discussing the rather complicated model suitable for $D=11$ supergravity, we will illustrate this procedure in two simpler models.

\begin{subsubsection}{$\mathbf{G=Gl(d)}$}\label{Gl}
For the group of invertible $d\times d$ matrices $Gl(d)$, we choose the minimal non-trivial representation for the coordinates that is $\R^d$. If we use the Lorentz group $SO(d-1,1)$ instead of the compact subgroup $K$, the coset is without loss of generality \cite{Hi08} parametrized by (\ref{Coset})
\be\label{CosetGL}
\cV &\in&  Gl(d)/SO(d-1,1).
\ee
The Maurer--Cartan form (\ref{MCf}) again decomposes into
\be\label{MCf2}
\cV^{-1}\cdot d\cV &=:& \left(\cP_{\cza} + \cQ_{\cza}\right){\mathbf{R}(\cV)_{\cm}}^{\cza}\, dx^{\cm}\,\in\,\mathfrak{gl}_d
\ee
with the one-form valued Lie algebra elements $\cP\in \mathfrak{gl}_d\ominus \mathfrak{so}_{(d-1,1)}$, $\cQ\in \mathfrak{so}_{(d-1,1)}$ and $\cm,\cza=0,\dots,d-1$.\\

In order not to overburden the notation, the matrix representation $\mathbf{R}$ of the $(\mathfrak{g}=\mathfrak{gl}_d)$-elements $\cP$ and $\cQ$ will be simply denoted by adding indices $\cza,\czb,\ldots=0,\ldots,d-1$ to $\cP$ and $\cQ$. Then, it is an immediate corollary of the theorem 3.2 in \cite{Hi08} that the general action $S$, being at most quadratic in derivatives, for this coset (\ref{CosetGL}) is of the form
\be\label{Action4}
S&=&\int\limits_{\R^{d}}\det(\cV)d^{d}x\,
\Big(
r_0\eta^{\cza\czb}\underline{\D}_{\cza}{\left(\cP_{\czb}\right)_{\czc}}^{\czc}
+r_1\underline{\D}_{\czc}{\left(\cP_{\cza}\right)_{}}^{\cza \czc}
\\
&&
+r_2{\left( \cP_{\cza}\right)^{\cza \czc}} {\left( \cP_{\czd}\right)_{\czc}}^{\czd}
 +r_3{\left( \cP_{\czd}\right)^{\cza \czc}} {\left( \cP_{\cza}\right)_{\czc}}^{\czd}
+r_4\eta^{\cza\czb}{\left( \cP_{\cza}\right)_{\czd}}^{\czc}  {\left(\cP_{\czb}\right)_{\czc}}^{\czd} 
\nn\\
&&
+r_5{\left(\cP_{\cza}\right)_{}}^{\cza \czc}{\left(\cP_{\czc}\right)_{\czd}}^{\czd}
 +r_6\eta^{\czb\czc}{\left(\cP_{\czb}\right)_{\cza}}^{\cza}{\left(\cP_{\czc}\right)_{\czd}}^{\czd}
\Big) \nn
\ee
with $r_0,\ldots,r_6\in \R$, the invariant tensor of $SO(d-1,1)$ or Minkowski metric $\eta$ and the covariant derivative $\D$ (\ref{covder}) acting on the $SO(d-1,1)$-tensor $\cP$:
\beg
\underline{\D}_{\cza}{\left(\cP_{\czb}\right)_{\czc}}^{\czd} &=&\underline{\p}_a{\left(\cP_b\right)_{c}}^{d}
+{\left(\cQ_{\cza}\right)_{\czb}}^{\cze}{\left(\cP_{\cze}\right)_{\czc}}^{\czd}
+{\left(\cQ_{\cza}\right)_{\czc}}^{\cze}{\left(\cP_{\czb}\right)_{\cze}}^{\czd}
-{\left(\cQ_{\cza}\right)_{\cze}}^{\czd}{\left(\cP_{\czb}\right)_{\czc}}^{\cze}.
\eeg
This is a seven parameter family of actions with manifest $Gl(d)$ invariance.\footnote{This could be reduced to a five-parameter family by isolating total derivative terms. In order to match the Einstein--Hilbert action in the form (\ref{EH}), we dispense with this simplification.} There is one particular choice
\be\label{Param4}
r_0\,=\,-\frac12,\quad r_1\,=\,\frac12, && r_2\,=\,-\frac12,\quad r_3\,=\,\frac12,\\
r_4\,=\,-\frac14,\quad r_5\,=\,\frac12, && r_6\,=\,-\frac14,\nn
\ee
with the property that the symmetry $Gl(d)$ is enlarged to $\Diff(d)$. This immediately follows from the fact that the action $S$ (\ref{Action4}) with the constants (\ref{Param4}) is a different way of writing the Einstein--Hilbert action $S_{EH}$ (\ref{EH}).\\

The independent degrees of freedom of the coset element $\cV$ (\ref{CosetGL}) exactly match the ones of the symmetric $G$-tensor
\be\label{Metrik}
g_{\cm\cn} &=& \mathbf{R}(\cV)_{\cm}{}^{\cza} \mathbf{R}(\cV)_{\cn}{}^{\czb} \eta_{\cza\czb}.
\ee
Note that this notation allows to rewrite the entire formalism of generalized coset dynamics in the $G$-covariant frame by substituting 
\beg
(\cP_{\czc})_{\cza\czb}
&=&
\frac12 \mathbf{R}(\cV^{-1})_{\cza}{}^{\cm}\mathbf{R}(\cV^{-1})_{\czb}{}^{\cn}\,\underline{\p}_{\czc} g_{\cm\cn}
\eeg
in the Lagrangian (\ref{Action4}). For the choice of constants (\ref{Param4}), it is obvious that the symmetric $G$-tensor $g$ can be identified with the metric $g$ (\ref{VielbeinG}) what justifies using the same symbol for both objects. We will stick to the $K$-covariant frame however, because it allows to discuss the induced action on representations of the covering group of $K\subset G$, on which there is no direct action of $G$ in general. \newpage

One might argue that $G=Gl(d)$ is a quite trivial example, because the coset element $\cV\in G/K$ merely is the vielbein introduced in (\ref{VielbeinG}), but it provides the general idea. This interpretation of $\cV$ as the vielbein allows to replace the domain of integration $\R^d$ in the action (\ref{Action4}) by an unrestricted manifold $\cM^d$ in the usual sense for the choice of constants (\ref{Param4}).
\end{subsubsection}

\begin{subsubsection}{$\mathbf{G=Sp(2n)}$}\label{Sp}
In a next step towards supergravity, we would like to consider the symplectic Lie group $G=Sp(2n)$ with compact subgroup $K=U(n)$.\footnote{In the matrix representation of $Sp(2n)$ as $2n\times 2n$ matrices, the subgroup of antisymmetric matrices forms a representation of $U(n)$ \cite{Hi08}.} Since the matrix representation of the group $Sp(2n)$ is of unit determinant and since $Sp(2n)$ is a subgroup of $Gl(2n)$, the general action $S$ being at most quadratic in derivatives, is immediately deduced from the action $S$ of the $G=Gl(d)$ case (\ref{Action4}) with indices $\cza,\czb,\ldots=1,\dots,2n$:
\be\label{Action5}
S&=&\int\limits_{\R^{2n}}d^{2n}x\,
\Big(
r_1\underline{\D}_{\czc}{\left(\cP_{\cza}\right)_{}}^{\cza \czc}
\\
&&
+r_2{\left( \cP_{\cza}\right)^{\cza \czc}} {\left( \cP_{\czd}\right)_{\czc}}^{\czd}
 +r_3{\left( \cP_{\czd}\right)^{\cza \czc}} {\left( \cP_{\cza}\right)_{\czc}}^{\czd}
+r_4\eta^{\cza\czb}{\left( \cP_{\cza}\right)_{\czd}}^{\czc}  {\left(\cP_{\czb}\right)_{\czc}}^{\czd} 
\Big). \nn
\ee
Hence, there only is a four-parameter family of actions for the case of a symplectic coset element. Which symmetry enlargement can we expect in this case?\\

It is obvious that the coset element $\cV\in Sp(2n)$ cannot be identified with a general vielbein any more and hence $\Diff(2n)$ symmetry is ruled out. However, the lower dimensional symmetry $\Diff(n)$ can be obtained for specific choices of $r_1,r_2,r_3$. This can be seen as follows.\\

Assume that a theory is $\Diff(m)$-invariant for some  $m\in \N$. Then the subgroup $Gl(m)\subset \Diff(m)$ with constant Jacobian matrix $\frac{\p\vp}{\p x}$ must be a symmetry. The transformation (\ref{CosTrafo2}) implies that this $Gl(m)$ further has to be a subgroup of $Sp(2n)$. Therefore, a necessary condition for $\Diff(m)$ symmetry is that $Gl(m)$ is a subgroup of $Sp(2n)$. The solution to this representation theoretical problem obviously leads to $m\leq n$ \cite{Hi08}.\\

The proof that the maximal diffeomorphism symmetry $\Diff(n)$ can be realized, is not too complicated. Start with the well-known observation that $\Diff(n)$ is a subgroup of the group of symplectomorphisms $\Symp(2n)$. This is the subgroup of $\Diff(2n)$ that preserves a non-degenerate symplectic form. On the other hand, it is clear that the action $S$ (\ref{Action4}) with the constants (\ref{Param4}) has the symmetry group $\Diff(2n)$ for $2n=d$. Hence, it is in particular invariant under its subgroup $\Symp(2n)$. Furthermore, the latter symmetry respects the choice $\cV\in Sp(2d)$. Therefore, the constants in the action $S$ (\ref{Action5}) to obtain $\Diff(n)$-invariance have to be the same as in (\ref{Param4}): 
\be\label{Param5}
r_1\,=\,\frac12,\quad r_2\,=\,-\frac12,&&r_3 \,=\, \frac12,\quad r_4=-\frac14.
\ee

The interpretation of the coset $\cV\in Sp(2n)/U(n)$ as a vielbein on a manifold would only be consistent, if the transition matrices of coordinate charts were also in $Sp(2n)$. This is the case for a symplectic manifold $(\cM^{2n},\Omega)$, a $2n$ dimensional manifold $\cM^{2n}$ with a non-degenerate closed symplectic form $\Omega$. Darboux's theorem guarantees the existence of an atlas of coordinate charts whose transition matrices are symplectic and in which the symplectic form has constant canonical form \cite{McDS95}. Hence, it is possible to replace the domain of integration $\R^{2n}$ in the action $S$ (\ref{Action5}) for the choice of constants (\ref{Param5}) by any symplectic manifold $(\cM^{2n},\Omega)$.\\

This is a first example of a constrained geometry: It is consistent to restrict the degrees of freedom of the vielbein on a symplectic manifold with a metric of Euclidean signature to $Sp(2n)/U(n)$.\footnote{Keeping in mind that the names of the indices are arbitrary, the statement $\cV\in Sp(2n)$ is equivalent to $\mathbf{R}(\cV)_{\cm}{}^{\cza} = \Omega_{\cm\cn}\mathbf{R}(\cV^{-1})_{\czb}{}^{\cn}\Omega^{\czb\cza}$ with the canonical symplectic form $\Omega_{\cza\czb}$ and $\Omega^{\cza\czb}\Omega_{\czb\czc}:=\delta_{\czc}^{\czb}$. The definition of the metric $g_{\cm\cn} = \mathbf{R}(\cV)_{\cm}{}^{\cza} \mathbf{R}(\cV)_{\cn}{}^{\czb} \eta_{\cza\czb}$ (\ref{Metrik}) is however equivalent to $\mathbf{R}(\cV)_{\cm}{}^{\cza} = g_{\cm\cn}\mathbf{R}(\cV^{-1})_{\czb}{}^{\cn}\eta^{\czb\cza}$. More details and further references can be found in \cite{Hi08}.
}
\end{subsubsection}

\begin{subsubsection}{$\mathbf{G=\big(Gl(4)\times E_{7(7)}\big)\ltimes \cN_{(4,\,56)}}$}
\label{GE7}
The group that will prove to be relevant for $D=11$ supergravity is $G=\big(Gl(4)\times E_{7(7)}\big)\ltimes \cN_{(4,\,56)}$, the semi-direct product of the product of the classical groups $Gl(4)$ and $E_{7(7)}$ with the nilpotent group $\cN_{(4,\,56)}$, which is the tensor product of the lowest possible representations $\mathbf{4}$ of $Gl(4)$ and $\mathbf{56}$ of $E_{7(7)}$. $G$ is best understood by its matrix representation as $60\times 60$ matrices:
		\be\label{GenV}
	\cV
	&\in&
	\left(\begin{tabular}{c|c}
	$Gl(4)
	$ & $*_{\text{\tiny4x56}}$\\
	\hline
	$0$												& $E_{7(7)}$
	\end{tabular}
	\right).
	\ee
This structure is obviously tailored for a coset construction $\cV\in G/K$ with the de Wit--Nicolai group 
\beg
K =SO(3,1)\times SU(8)/\Z_2.
\eeg
In particular, any Lagrangian necessarily has the local covariance group $SO(3,1)\times SU(8)/\Z_2$ of $D=11$ supergravity, if it is constructed from $\cV$ (\ref{GenV}) using the generalized coset dynamics. We have shown in section \ref{KK} that the Lorentz covariance group is not affected by a dimensional reduction a priori. Hence, reducing from the sixty dimensional setting to $D=11$ does not affect the $SO(3,1)\times SU(8)/\Z_2$ covariance. A further reduction to $d=4$ trivially provides a theory with global manifest $E_{7(7)}$-invariance.\\

To prove that there is an action with $60$ coordinates that reduces to $D=11$ supergravity amounts to showing that the general class of Lagrangians with $G$-invariance contains one with the particular property that the diffeomorphism symmetry of $D=11$ supergravity is restored, at least for a subset of solutions with $49$ independent Killing vectors.\\

The last subclause is essential: For the cases $G=Gl(d)$ and $G=Sp(2n)$, there existed extensions to infinite dimensional groups $\Diff(d)$ and $\Symp(2n)$ respectively. Cartan's theorem \cite{C09} implies that this can be ruled out for $E_{7(7)}$, because there is no infinite dimensional subgroup $H$ of $\Diff(56)$ containing $E_{7(7)}$ with the property that for every $\vp\in H$, the Jacobian matrix $\frac{\p \vp}{\p x}$ is an $E_{7(7)}$ element. Hence, $G$ cannot be extended to a symmetry group that contains $\Diff(4)\times\Diff(7)\subset \Diff(11)$ without violating $\cV\in G/K$ \cite{Hi08}.\\

The crucial observation is that we do not have to require this at all. Since $D=11$ supergravity does not know anything about the remaining $49$ coordinates, it is completely sufficient that a subset of solutions with $49$ independent Killing vectors forms an orbit of $\Diff(11)$.\\

This is to be understood in complete analogy to the space of solutions of the wave equation in $d=4$ dimensional flat Minkowski space. The set of solutions depending on all spacetime dimensions forms an orbit of the finite dimensional conformal symmetry group $SO(4,2)/\Z_2$. The subspace of solutions that depend on two spacetime dimensions however forms an orbit of the infinite dimensional symmetry group of conformal transformations in two dimensions. \\

In this article, we will not construct the complete generalized coset dynamics in sixty dimensions that is expected to have a hidden $\Diff(11)$ symmetry in its reduction to $D=11$. We will content ourselves with the $56$-dimensional sector that corresponds to the $E_{7(7)}$ part of the group $G$ (\ref{GenV}).\\ 

In the next section, we will prove that there is a Lagrangian in $56$ dimensions that exactly reproduces the dynamics of $D=11$ supergravity upon dimensional reduction to the seven common dimensions, if only the degrees of freedom that are encoded in the $E_{7(7)}$-valued $56\times 56$ submatrix of $\cV$ (\ref{GenV}) are taken into account. Furthermore, we will explain with mere group theoretical arguments in section \ref{restdiff7} why adding four additional dimensions appears to be preferred.
\end{subsubsection}
\end{subsection}
\end{section}

\begin{section}{Bosonic dynamics}
\label{BosDyn}
\begin{subsection}{$\mathbf{G=E_{7(7)}}$}
\label{E7Action}
The discussion in the previous section indicates that the investigation of the generalized coset dynamics for the Lie group $G=E_{7(7)}$ may be interesting for supergravity. Hence, we focus on the coset
\be\label{cosetE7}
\cV&\in&  E_{7(7)}/(SU(8)/\Z_2).
\ee
The lowest dimensional, non-trivial representation space of $E_{7(7)}$ is $\R^{56}$, on which the group acts as prescribed in (\ref{xTrafo}). Since $E_{7(7)}$ also preserves a symplectic form, it is a subgroup of $Sp(56)$. Hence, the action (\ref{Action5}) from section \ref{Sp} provides the general ansatz to construct the dynamics. The general action with $E_{7(7)}$-invariance that is of at most second order in derivatives and exclusively depends on the coset element $\cV$ (\ref{cosetE7}) reads
\be\label{Action6}
S&=&\int\limits_{\R^{56}}d^{56}x\,
\Big(
r_1\underline{\D}_{\czc}{\left(\cP_{\cza}\right)_{}}^{\cza \czc}
\\
&&
+r_2{\left( \cP_{\cza}\right)^{\cza \czc}} {\left( \cP_{\czd}\right)_{\czc}}^{\czd}
 +r_3{\left( \cP_{\czd}\right)^{\cza \czc}} {\left( \cP_{\cza}\right)_{\czc}}^{\czd}
+r_4\eta^{\cza\czb}{\left( \cP_{\cza}\right)_{\czd}}^{\czc}  {\left(\cP_{\czb}\right)_{\czc}}^{\czd} 
\Big). \nn
\ee
The objects $\cP$ and $\cQ$ follow the definition (\ref{MCf2}). The matrix representation of $E_{7(7)}$ as $56\times 56$-matrices provides the canonical embedding of $E_{7(7)}$ in $Gl(56)$. Therefore, the indices $\cza,\czc,\dots$ in the action $S$ (\ref{Action6}) take the values $1,\dots,56$. \\

The maximal compact subgroup of $E_{7(7)}$ is $K=SU(8)/\Z_2$ (\ref{cosetE7}). In the matrix representation of $E_{7(7)}$ as $56\times 56$ matrices, the corresponding compact group elements are presented as real orthogonal matrices due to the embedding $SU(8)/\Z_2\subset SO(56)$. This implies that it does not matter if the indices $\cza,\czc,\dots$ in the action (\ref{Action6}) are raised or lowered, because their position can be freely adjusted with the symmetric invariant tensor $\eta_{\cza\czb}$ of $SO(56)$ in complete analogy to the $Gl(d)$-case from section \ref{Gl}.\footnote{Since $SU(8)/\Z_2$ also is a subgroup of $Sp(56)$, it is of course equivalently possible to raise and lower the $K$-indices $\cza,\czb,\dots$ with the symplectic form $\Omega$.
}\\

In a further step, we adapt the notation to the symmetry structure. This is achieved by observing that the $56$ dimensional representation space of $E_{7(7)}$ splits into two irreducible representations of $SU(8)/\Z_2$:
\be\label{holom13}
\mathbf{56} &=& \mathbf{28} + \overline{\mathbf{28}}.
\ee
It is important to note that the real Lie group $SU(8)/\Z_2$ necessitates a complex $\mathbf{28}$ dimensional vector space to act on. Hence, it is natural to introduce $28$ holomorphic coordinates instead of $56$ real ones. The contragredient or dual representation $\overline{\mathbf{28}}$ in (\ref{holom13}) then simply corresponds to the antiholomorphic coordinates, i.e. a complex conjugation. Since the group $K$ is $SU(8)/\Z_2$ and not $U(28)$, it is furthermore appropriate to label these holomorphic coordinates by the antisymmetric pair $[AB]$ with $A,B=1,\dots,8$. Hence, the one-form $\cP$ (\ref{MCf2}) decomposes into
\be\label{cPdefi2}
\cP &=:&
\cP_{AB}\mathbf{R}(\cV){}_{\cm}{}^{AB}dx^{\cm} +\text{c.c.}
\ee
with $\cm=1,\dots,56$ and the abbreviation c.c. for complex conjugation. In contrast to the $SO(56)$ indices $\cza$, the position of the $SU(8)/\Z_2$ indices $AB$ is not arbitrary: Lowering or raising indices is equivalent to a complex conjugation. We make use of the standard convention to distinguish complex conjugated objects only by the position of their $SU$-indices, e.g. $\cP^{AB}=\cP_{AB}^*$. In this notation, equation (\ref{cPdefi2}) reads
\beg
\cP &=:& \left(\cP_{AB}\mathbf{R}(\cV){}_{\cm}{}^{AB} +\cP^{AB}
\mathbf{R}(\cV){}_{\cm,AB}\right)dx^{\cm}
\eeg

Relabelling the indices, the action (\ref{Action6}) takes the form
\be\label{Action7}
S&=&\int\limits_{\R^{56}}d^{56}x\,
\Big(
r_1\left(\underline{\D}_{AB}{\left(\cP_{CD}\right)_{}}^{ABCD} +\text{c.c.}\right)
+r_2\left( \cP_{AB}\right)^{ABCD} \left( \cP^{EF}\right)_{CDEF}
\nn\\
&&
 %
 +r_3\left( \cP_{AF}\right)^{ABCD} \left( \cP^{EF}\right)_{EBCD}
+r_4\left( \cP_{EF}\right)^{ABCD}  \left(\cP^{EF}\right)_{ABCD}
\Big). 
\ee
The strong restriction $\cV\in E_{7(7)}/(SU(8)/\Z_2)$ is the reason why not more terms appear in this expansion. The one forms $\cP\in \mathfrak{e}_{7(7)}\ominus \mathfrak{su}_{8}$ and $\cQ\in \mathfrak{su}_{8}$ form the irreducible $\binom{8}{4}=\mathbf{70}$ and $\mathbf{63}$ dimensional $\mathfrak{su}_{8}$-representation spaces respectively. The $\e$-tensor in eight dimensions links the one form $\cP_{EFGH}$ to its complex conjugated $\cP^{EFGH}=\cP_{EFGH}^*$ in an $\mathfrak{su}_{8}$-covariant way:
\be\label{dual}
\cP^{ABCD}
&=& \frac{1}{4!}\e^{ABCDEFGH}\cP_{EFGH}.
\ee
We refrain from calling the one-form $\cP$ with four completely antisymmetrized $\mathfrak{su}_{8}$-indices ``selfdual'', because the $\e$ tensor relates complex conjugated objects in this case. To conclude this section, we remark that the complex conjugate only has to be added to the first term in the action $S$ (\ref{Action7}). Due to the relation (\ref{dual}), the other three contributions are real on their own.
\end{subsection}

\begin{subsection}{The hidden symmetry $\Diff(7)$}
\label{DecomGl7}
As explained in section \ref{GE7}, we will fix the constants $r_1,\dots,r_4\in \R$ in the action $S$ (\ref{Action7}) such that some diffeomorphism symmetry is restored. Due to Cartan's theorem \cite{C09}, this can only be possible for a subset of solutions with $49$ independent Killing vectors or equivalently, in a Kaluza--Klein reduction.\\

In complete analogy to the argument for the case of $G=Sp(2n)$ in section \ref{Sp}, a necessary criterion for $\Diff(m)$ to be a symmetry group is that $Gl(m)$ be a subgroup of $E_{7(7)}$. The maximal solution to this representation theoretical problem is $Gl(7)\subset E_{7(7)}$. Therefore, we will try to choose $r_1,\dots,r_4\in \R$ such that $\Diff(7)$ is a hidden symmetry of the action $S$ (\ref{Action7}). To achieve this, it is natural to first parametrize the Lie algebra $\mathfrak{e}_{7(7)}$ by $\mathfrak{gl}_7$-representations:
\begin{center}
\begin{tabular}{ccccccc}
$\mathbf{133}$ &$=$&$\mathbf{49}$&$\oplus$ &$\big(\mathbf{35}\oplus\overline{\mathbf{35}}\big)$&$\oplus$&$\big(\mathbf{7}\oplus\overline{\mathbf{7}}\big)$.
\end{tabular}
\end{center}

As a next step, recall from section \ref{Gl} that the Lagrangian can be written purely in terms of the symmetric $G$-tensor $g$ (\ref{Metrik}). Hence, it is obvious that only the $\mathbf{133}-\mathbf{63}=\mathbf{70}$ degrees of freedom of the completely gauge fixed coset element $\cV\in E_{7(7)}/(SU(8)/\Z_2)$ appear in the action $S$ (\ref{Action7}). Therefore, we can without loss of generality partly fix the $SU(8)/\Z_2$-symmetry to $SO(7)$ by requiring that the coset follows the block-triangular decomposition
\be
\cV&=:&e^{{h_a}^b{\left.\hat{M}\right.^a}_b}
e^{A_{abc}\hat{E}^{abc}}e^{A_{a_1\dots a_6}\hat{E}^{a_1\dots a_6}}
\label{cosetE73}
\ee
with the indices $a,b,\ldots=4,\dots,10$, the matrix exponential $e$ and the $\mathfrak{e}_{7(7)}$-generators $\hat{M}$ and $\hat{E}$ in their representation as $56\times 56$ matrices, whose non-vanishing commutation relations are \cite{Hi08}
\begin{subequations}\label{M2E}
\be
 \left[{\left.\hat{M}\right.^{e}}_f, {\left.\hat{M}\right.^{g}}_h\right] &=& \delta_f^g {\left.\hat{M}\right.^{e}}_h -\delta_h^e {\left.\hat{M}\right.^{g}}_f
   \label{ComRel1}
   \\
	\left[ {\left.\hat{M}\right.^{e}}_f,\hat{E}^{abc}\right]\label{E2E3}
  &=&
 3\delta_f^{[a}\hat{E}^{bc]e}
   \\
  \label{E2E6}
\left[ {\left.\hat{M}\right.^{a}}_b,\hat{E}^{e_1\dots e_6}\right]
	&=&6\delta_b^{[e_6}\hat{E}^{e_1\dots e_5]a}
		\\
	\left[ \hat{E}^{efg},\hat{E}^{abc}\right]\label{E3E3}
	 &=&40 \hat{E}^{efgabc}.
\ee
\end{subequations}
Since we have not fixed the $SO(7)$ in the $SU(8)/\Z_2$ symmetry, the $Gl(7)$ part in (\ref{cosetE73}) in fact is a coset
\be\label{gl1}
e^{{h_a}^b{\left.\hat{M}\right.^a}_b}&\in&Gl(7)/SO(7).
\ee
As soon as the $\Diff(7)$ symmetry is established, the $Gl(7)/SO(7)$ will be parametrized by an unrestricted vielbein in seven dimensional Euclidean space.
\newpage

The decomposition of $E_{7(7)}$ into $Gl(7)$ representations also uniquely induces a decomposition of the $\mathbf{56}$ dimensional representation space into $Gl(7)$ representation spaces:
\be\label{egl7}
\mathbf{56} &=& \mathbf{7}\oplus\overline{\mathbf{21}}\oplus\mathbf{21}\oplus\overline{\mathbf{7}}.
\ee
Contragredient representations of $Gl(7)$ appear in this decomposition of $\mathbf{56}$. This is of course expected due to the fact that $E_{7(7)}$ is a subgroup of $Sp(56)$. Hence, there is a preserved symplectic structure $\Omega$ which is the reason why we will denote the variables in dual representations by momenta $p$ instead of coordinates $x$. Therefore, one can without loss of generality arrange the labelling of the $56$ coordinates so as to make manifest the decomposition (\ref{egl7}) namely
\begin{center}
\begin{tabular}{cccccc}
$\frac{\p}{\p x^{m}}$&$:=$&$\delta^{\cm}_{m-3}\frac{\p}{\p x^{\cm}},$&\quad
$\frac{\p}{\p p_{mn}}$&$:=$&$\delta^{\cm-7,\,[mn]}\frac{\p}{\p x^{\cm}},$\\
$\frac{\p}{\p x^{mn}}$&$:=$&$\delta^{\cm-28}_{[mn]}\frac{\p}{\p x^{\cm}},$&\quad
$\frac{\p}{\p p_m}$&$:=$&$\delta^{\cm-49,\,m-3}\frac{\p}{\p x^{\cm}}$
\end{tabular}
\end{center}
with the range of the indices $\cm=1,\dots,56$, $m,n=4,\dots,10$ and $\delta^{1}_{[4\,5]}=1$ et cetera. We will use the same labelling for the derivatives $\underline{\p}$ (\ref{underl})
\begin{center}
\begin{tabular}{cccccc}
$\underline{\p}_a$&$:=$&$\delta^{\cza}_{a-3}\underline{\p}_{\cza},$&\quad
$\underline{\p}^{ab}$&$:=$&$\delta^{\cza-7,\,[ab]}\underline{\p}_{\cza},$\\
$\underline{\p}_{ab}$&$:=$&$\delta^{\cza-28}_{[ab]}\underline{\p}_{\cza},$&\quad
$\underline{\p}^{a}$&$:=$&$\delta^{\cza-49,\,a-3}\underline{\p}_{\cza}$
\end{tabular}
\end{center}
with $\cza=1,\dots,56$ and $a,b=4,\dots,10$ respectively. Then, the relation $ \frac{\p}{\p x^{\cm}}= \mathbf{R}(\cV){{}_{\cm}}^{\cza}\,\underline{\p}_{\cza}$ (\ref{underl}) can be written in the matrix formalism with the obvious contraction of indices:
\beg
\left(
\scalebox{0.8}{
\begin{tabular}{c}
$\frac{\p}{\p x^m}$\\
$\frac{\p}{\p p_{mn}}$\\
$\frac{\p}{\p x^{mn}}$\\
$\frac{\p}{\p p_m}$
\end{tabular}
}\right)
&\scalebox{0.8}{=}&
\left(
\scalebox{0.8}{
\begin{tabular}{l|l|l|l}
$\mathbf{R}(\cV)_m{}^a$ & ${\mathbf{R}(\cV)_{m,ab}}_{}$ & $\mathbf{R}(\cV)_{m}{}^{ab}$ & ${\mathbf{R}(\cV)_{m,a}}_{}$\\
\hline
${\mathbf{R}(\cV)^{mn,a}}^{}$ & $\mathbf{R}(\cV)^{mn}{}_{ab}$ & ${\mathbf{R}(\cV)^{mn,ab}}^{}$ & $\mathbf{R}(\cV)^{mn}{}_{a}$\\
\hline
$\mathbf{R}(\cV)_{mn}{}^{a}$ & ${\mathbf{R}(\cV)_{mn,ab}}_{}$ & $\mathbf{R}(\cV)_{mn}{}^{ab}$ & ${\mathbf{R}(\cV)_{mn,a}}_{}$\\
\hline
${\mathbf{R}(\cV)^{m,a}}^{}$ & $\mathbf{R}(\cV)^{m}{}_{ab}$ & ${\mathbf{R}(\cV)^{m,ab}}^{}$ & $\mathbf{R}(\cV)^{m}{}_{a}$
\end{tabular}
}\right)
\left(\scalebox{0.8}{
\begin{tabular}{c}
$\underline{\p}_a$\\
$\underline{\p}^{ab}$\\
$\underline{\p}_{ab}$\\
$\underline{\p}^{a}$
\end{tabular}
}
\right)
\eeg
It follows from the commutation relations (\ref{M2E}) that the parametrization of the coset $\cV$ (\ref{cosetE73}) really is block-triangular. This is equivalent to stating that the generators $\hat{E}$ are represented by nilpotent upper triangular $56\times 56$ matrices. Hence, the top left corner of the matrix representation of $\cV$ may only depend on the $Gl(7)/SO(7)$-degrees of freedom (\ref{gl1}). Therefore, these can be parametrized by
\be\label{eDefi}
\mathbf{R}(\cV)_m{}^a &=:& \Delta^{\frac12}{e_m}^a
\ee
with $\Delta:=\det({e_m}^a)$ as in (\ref{sigmaDefi2}). It should be noted that the $Gl(7)$ embedding in $E_{7(7)}$ is unique only modulo a rescaling by the $Gl(1)$ factor corresponding to $\Delta$ \cite{Hi08}. The choice (\ref{eDefi}) will prove appropriate to uncover the hidden $\Diff(7)$ symmetry. \newpage

With this definition and the convention to denote the inverse ``siebenbein'' ${e_a}^m$ simply by a different naming of indices, i.e. ${e_m}^a{e_a}^n=\delta_m^n$, the gauged fixed coset element $\cV\in E_{7(7)}/(SU(8)/\Z_2)$ is represented as a $56\times 56$ matrix in the following way \cite{Hi08}
\beg
\mathbf{R}(\cV)=\left(
\scalebox{.75}{
\begin{tabular}{c|c|c|c}
$\Delta^{\frac12}{e_m}^a$
&$-\sqrt{2}\Delta^{\frac12}{e_m}^c A_{abc}$
&$\sqrt{2}\Delta^{\frac12}{e_m}^c {U_{c}}_{+}^{ab}$
&$-\Delta^{\frac12}{e_m}^c U_{ac}$\\
\hline
$0$&$\Delta^{\frac12}
 {e_a}^m{e_b}^n$
 &$-\frac{1}{6}\Delta^{\frac12}
 {e_c}^m{e_d}^n
 A_{a_1\dots a_3}\e^{cda_1\dots a_3ab}$
 &$\sqrt{2}\Delta^{\frac12}
 {e_c}^m{e_d}^n
 {U_{a}}_{-}^{cd}$\\
\hline
$0$&$0$&$\Delta^{-\frac12}
 {e_m}^a {e_n}^b$
 &$-\sqrt{2}\Delta^{-\frac12}
 {e_m}^c {e_n}^d A_{cd a}$\\
\hline
$0$&$0$&$0$&$\Delta^{-\frac12}{e_a}^m$
\end{tabular}
}
\right)
\eeg
with the abbreviations
\begin{subequations}\label{UDefi}
\be
{U_{d}}_{-}^{jk}
&:=&\frac{1}{12}A_{abc}A_{ghd}\e^{jkghabc}
-\frac{1}{360}A_{a_1\dots a_6}\e^{a_1\dots a_6[j}\delta_d^{k]},\\
{U_{d}}_{+}^{jk}
&:=&\frac{1}{12}A_{abc}A_{ghd}\e^{jkghabc}
+\frac{1}{360}A_{a_1\dots a_6}\e^{a_1\dots a_6[j}\delta_d^{k]},\\
U_{ad}
&:=&
\frac{1}{180}A_{a_1\dots a_6}A_{abd}\e^{ba_1\dots a_6}
+\frac{1}{18} A_{ars}A_{ghi}A_{kld}\e^{rsghikl}.
\ee
\end{subequations}
\end{subsection}

\begin{subsection}{Connecting $\mathbf{Gl(7)}$- and $\mathbf{SU(8)/\Z_2}$-representations}
\label{SU8Gl7}
In order to be able to decide, whether there are constants $r_1,\dots,r_4\in \R$ for the action $S$ (\ref{Action7}) such that a hidden $\Diff(7)$ symmetry appears in a Kaluza--Klein reduction, the $Gl(7)$ representations of section \ref{DecomGl7} have to be linked to the $SU(8)/\Z_2$ representations that were used in the action $S$ (\ref{Action7}). This is done by first splitting the $Gl(7)$ representations into $SO(7)$ representations and recombine them afterwards into $SU(8)/\Z_2$ representations. For the last step, one has to introduce the Clifford algebra with the seven dimensional Euclidean metric $\eta$ and $a,b=4,\dots,10$
\be\label{Clifford}
\{\G^a,\G^b\} &=& 2\eta^{ab}.
\ee
It is well known that it has a representation in terms of purely imaginary matrices $\G^a\in i\R^{8\times8}$.\footnote{Furthermore, we define $\G^{a_1\dots a_n}:= \G^{[a_1}\cdots \G^{a_n]}$ with antisymmetrization of strength one and we fix the normalization $\G^{a_1\dots a_7}=-i\e^{a_1\dots a_7}\id_{8}$ with $\e^{1\,2\,3\,4\,5\,6\,7}=1$.}\\

The transformation between the holomorphic cooordinate frame defined by the decomposition (\ref{holom13}) and the one of $Gl(7)$ (\ref{egl7}) is provided by the following identification, e.g. for the derivative $\underline{\p}$ (\ref{underl}) with  $a,b=4,\ldots,10$ and $A,B=1,\ldots,8$ \cite{Hi08}
\be\label{defiSo7P2}
\underline{\p}_{AB} 
&=&
6i{\G^a}_{AB}
\left(\underline{\p}_{a} -i\eta_{ac} \underline{\p}^{c}\right)
-2\sqrt{2} {\G^{ab}}_{AB}
\left(\underline{\p}_{ab} -i\eta_{ac} \eta_{bd} \underline{\p}^{cd}\right).
\ee

With these definitions, it is a straightforward computation \cite{Hi08} to arrive at the following identifications for the components $\cP_\alpha$ and $\cQ_\alpha$ (\ref{MCf2}):
\begin{subequations}\label{vGL}
\be\label{vminE7}
{\left(\cQ_{\cza}\right)_{A}}^{B}
&:=&\frac{1}{3}\mathbf{R}(\cV^{-1})_{AC}{}^{\cm} \underline{\p}_{\cza} \mathbf{R}(\cV)_{\cm}{}^{BC}\nn\\
&=&
\frac{1}{4}{\left(\cQ_{\cza}\right)_{e}}^{f} {{{\G^e}_f}_A}^B
+\frac{1}{12}\left(\cP_{\cza}\right)_{a_1\dots a_3}{{\G^{a_1\dots a_3}}_A}^B
\nn\\
&&
-\frac{i}{1440}\left(\cP_{\cza}\right)_{a_1\dots a_6}\e^{a_1\dots a_6c}{{\G_{c}}_A}^B
\\
\label{vE7}
\left(\cP_{\cza}\right)^{ABCD}
&:=&\mathbf{R}(\cV^{-1})^{AB,\cm} \underline{\p}_{\cza} \mathbf{R}(\cV)_{\cm}{}^{CD}\nn\\
&=&
-\frac{3}{4}{\left(\cP_{\cza}\right)_{e}}^{f}
{\G^e}^{[AB}{\G_f}^{CD]}\nn\\
&&
-\frac{1}{4}\left(\cP_{\cza}\right)_{a_1\dots a_3}
{\G^{[a_1a_2}}^{[AB}{\G^{a_3]}}^{CD]}
\nn\\
&&
+\frac{i}{2880}\left(\cP_{\cza}\right)_{a_1\dots a_6}
\e^{a_1\dots a_6c}
{\G_{ec}}^{[AB}{\G^{e}}^{CD]}
\ee
\end{subequations}
with $\cm,\cza=1,\ldots,56$ and the abbreviations
\begin{subequations}\label{vSym}
\be
{\left(\cP_{\cza}\right)_{c}}^{d}&:=&
{e_{g}}^m \underline{\p}_{\cza} {e_m}^{(d}\eta^{f)g}\eta_{cf}
 \\
{\left(\cQ_{\cza}\right)_{c}}^{d}&:=&{e_{g}}^m \underline{\p}_{\cza} {e_m}^{[d}\eta^{f]g}\eta_{cf}
\\
\left(\cP_{\cza}\right)_{a_1\dots a_3}&:=&
\underline{\p}_{\cza} A_{a_1\dots a_3} 
+3A_{f[a_1a_2}{e_{a_3]}}^m \underline{\p}_{\cza} {e_m}^{f}
\\
\left(\cP_{\cza}\right)_{a_1\dots a_6}&:=& 
\underline{\p}_{\cza} A_{a_1\dots a_6}
-6 A_{f[a_1\dots a_5}{e_{a_6]}}^m \underline{\p}_{\cza} {e_m}^{f}
\\
&&
-20A_{[a_1\dots a_3}\left(\cP_{\cza}\right)_{a_4\dots a_6]}.\nn
\ee
\end{subequations}
The definitions of the first two objects in (\ref{vSym}) coincide with the ones (\ref{Abk}) that have already been used for supergravity in section \ref{DynSUSY}. The factor $\frac{1}{3}$ in (\ref{vminE7}) guarantees the standard normalization of the associated $\mathfrak{su}_8$ generators \cite{Hi08}.
\end{subsection}

\begin{subsection}{The bosonic Lagrangian}\label{Lagr}
This parametrization of $\cP$ and $\cQ$ in terms of $Gl(7)$ representations ${e_m}^a$, $A_{abc}$ and $A_{a_1\dots a_6}$ allows to decide whether the constants $r_1,\dots,r_4$ in the action $S$ (\ref{Action7}) can be chosen in such a way that the global $Gl(7)\subset E_{7(7)}$ symmetry of the space of solutions with $49$ independent Killing vectors can be enlarged to a local symmetry of the type of $\Diff(7)$. We find that there is a unique choice of $r_1,\dots,r_4$ leading to such an enlarged local symmetry:
\be\label{Action8}
\mathcal{L}&=&\left(\underline{\D}_{AB}{\left(\cP_{CD}\right)_{}}^{ABCD} +\text{c.c.}\right)
\\
&&
 %
 -\frac{8}{3}\left( \cP_{AF}\right)^{ABCD} \left( \cP^{EF}\right)_{EBCD}
+\frac{1}{6}\left( \cP_{EF}\right)^{ABCD}  \left(\cP^{EF}\right)_{ABCD}
. \nn
\ee
However, we only obtain a subgroup of $\Diff(7)$, namely the volume preserving diffeomorphisms, as a hidden symmetry. To prove this statement, we perform a Kaluza--Klein reduction of the Lagrangian (\ref{Action8}) to the seven coordinates that correspond to the $\mathbf{7}$ coordinates in the decomposition (\ref{egl7}), which leads to
\be\label{LagrKK}
\mathcal{L}_{KK}
&=&
2^7 9\Delta^{-1}\left(
\frac{1}{4}\tilde{R}_7
-\frac{1}{48}F_{b_1\dots b_4}F^{b_1\dots b_4}
-\frac{4!}{7!^2 48}\dF^2
\right)\\
&&+ 2^7 9 \left[  
    \frac{3}{4}\eta^{ab} \Delta^{-2}\p_a \Delta\, \, \Delta^{-1}\p_b \Delta
  +\frac{3}{4}\eta^{ab}\frac{\p}{\p x^{m}}\left(\Delta^{-1}{e_a}^m\, \Delta^{-1}\p_b \Delta \right)
  \right]\nn
\ee
with the Ricci scalar $\tilde{R}_7$ (\ref{Abk}) in $d=7$ and the abbreviations
\begin{subequations}\label{FDefi}
\be
  F_{abcd}&:=&4(\cP_{[a})_{bcd]}\\
  \dF&:=&\dF_{a_1\dots a_7}\e^{a_1\dots a_7}\\
&=&7(\cP_{a_1})_{a_2\dots a_7}\e^{a_1\dots a_7},\nn
    \ee
    \end{subequations}
    where we use the frame derivative $\p_a:={e_a}^m\frac{\p}{\p x^{m}}$ in (\ref{LagrKK}), in $\tilde{R}_7$ (\ref{Abk}) and in the definitions (\ref{vSym}) instead of $\underline{\p}_a$.\\
    
The parenthesis in the first line of the Lagrangian $\mathcal{L}_{KK}$ (\ref{LagrKK}) obviously is $\Diff(7)$ invariant, but the terms involving the determinant $\Delta$ do not match. (Note however that the last term is a total derivative.) A different parametrization of the $Gl(7)/SO(7)$-coset than (\ref{eDefi}), which would be equivalent to a Weyl rescaling in seven dimensions, does not cure this problem. Hence, one does not obtain $\Diff(7)$ as a hidden symmetry, but only its infinite dimensional subgroup of volume preserving diffeomorphisms that are characterized by a unimodular Jacobian matrix $\frac{\p \vp}{\p x}$ \cite{C09}. This problem can be cured by coupling additional dimensions to the setting, what we shall show next.\footnote{Furthermore, the Lagrangian $\mathcal{L}_{KK}$ (\ref{LagrKK}) reveals that the symmetries associated to the nilpotent generators $\hat{E}^{abc}$ and $\hat{E}^{a_1\dots a_6}$ (\ref{cosetE73}) (part of the Borel subgroup of $E_{7(7)}$) have been promoted to the two- and five-form gauge symmetries of supergravity.}
\end{subsection}

\begin{subsection}{Restoring $\Diff(7)$}\label{restdiff7}
Since the problematic terms in the Lagrangian $\mathcal{L}_{KK}$ (\ref{LagrKK}) only depend on the determinant $\Delta$, a Weyl rescaling of $d$ additional dimensions may provide the solution. Hence, one should investigate whether a Lagrangian in $7+d$ dimensions with symmetry group $\Diff(7+d)$ leads to the Lagrangian $\mathcal{L}_{KK}$ (\ref{LagrKK}) after a Weyl rescaling by $\Delta^z$ for $z\in \R$ of the vielbein that corresponds to the other $d$ dimensions. Fixing the $SO(6+d,1)$ symmetry to $SO(d-1,1)\times SO(7)$, we can without loss of generality assume the following shape for the vielbein $E$ in $7+d$ dimensions:
\be\label{Weyl2}
	E_{7+d}&=&
	\left(\begin{tabular}{c|c}
	$\Delta^{z}e_{\mu}{}^{\alpha}$ & $B_{\mu}{}^a$\\
	\hline
	$0$												& ${e_m}^a$
	\end{tabular}
	\right).
	\ee
The range of the indices is $\mu,\alpha=0,\dots,d-1$ and $m,a=d,\dots, d+6$. From the group theory point of view, this amounts to defining an action of the $Gl(1)[x]$ part of $\Diff(7)$ on the additional $d$ coordinates. Hence, the $d\times d$-part of the vielbein has to contain the corresponding factor of $\Delta$. Next, observe that the simplifying assumptions 
\begin{subequations}\label{assump}
\be
{e_\mu}^\alpha&=&\delta_\mu^\alpha\quad \text{and}\\
{B_\mu}^a&=&0
\ee
\end{subequations}
do not affect the contributions of the Ricci scalar $\tilde{R}_{7+d}$ that exclusively depend on $\Delta$. In this truncation, we obtain
\beg
{\det}(E_{7+d})\tilde{R}_{7+d} &=& \Delta^{zd+1}\left(\tilde{R}_{7} -zd[z(d+1)-4]\eta^{ab} \Delta^{-1}\p_a \Delta\, \, \Delta^{-1}\p_b \Delta\right)\\
&&-2zd\frac{\p}{\p x^m}\left(\Delta^{-1}{e_a}^m \Delta^{-1}\p_b\Delta\right)\eta^{ab}.
\eeg
The second line is a total derivative contribution, which does not alter the dynamics prescribed by the Lagrangian. Comparing this equation to the reduced Lagrangian $\mathcal{L}_{KK}$ (\ref{LagrKK}) leads to the unique solution
\begin{subequations}\label{values}
\be
d&=&4\\
z &=&-\frac{1}{2}.
\ee
\end{subequations}
This is a remarkable fact. The unique possibility to enlarge the $Gl(7)$ symmetry acting on the solutions of the action $S$ (\ref{Action8}) that only depend on the $\mathbf{7}$ coordinates in (\ref{egl7}) to $\Diff(7)$ is to consider a generalized coset that contains $d=4$ further directions, if one starts with Einstein--Hilbert actions in $7+d$ dimensions. The necessary Weyl rescaling by $\Delta^z$ is uniquely fixed by (\ref{values}), too. 
\end{subsection}

\begin{subsection}{Comparison to $\mathbf{D=11}$ supergravity}\label{CS1}
The conclusion of our study so far is that it appears to be ``natural'' from a pure $E_{7(7)}$ point of view to discuss a $56+4$ dimensional setting and the hidden symmetries in a truncation to $7+4$ dimensions. A comparison of the Weyl rescaling (\ref{values}) with (\ref{Vielbdecom}) furthermore shows that this exactly is the one used for $D=11$ supergravity to reveal the $E_{7(7)}$ symmetry in the truncation to four dimensions \cite{CJ79}. \newpage

Therefore, it looks promising to try to establish a link between the parametrization of the coset element $\cV$ (\ref{cosetE73}) and the fields of $D=11$ supergravity. For the $G=E_{7(7)}$ case discussed in this section, the Lagrangian $\mathcal{L}_{KK}$ (\ref{LagrKK}) only contains the siebenbein ${e_m}^a$, the four-form field strength $F_{b_1\dots b_4}$ and the seven-form field strength $\dF_{a_1\dots a_7}$. Since the relations (\ref{FDefi}) completely agree with the definitions of $F$ (\ref{F4a}) and of $\dF$ (\ref{7Form}) in supergravity, one can identify these objects as already anticipated by using the same notation. Substituting a four form field strength $F$ in four dimensions for the seven-form field strength $\dF$ in seven dimensions by the standard definition of supergravity (\ref{4Form})
\be\label{4Form2}
F^{\alpha_1\dots \alpha_4} &:=& \frac{1}{7!} \e^{\alpha_1\dots \alpha_4 a_1\dots a_7}\dF_{a_1\dots a_7},
\ee
the Lagrangian $\mathcal{L}_{KK}$ (\ref{LagrKK}) with the Weyl rescaling and the truncation (\ref{assump}) of section \ref{restdiff7} takes the form modulo total derivative terms and modulo a constant rescaling
\be\label{LagrKK2}
\mathcal{L}_{KK}
=
\det(E_{11})\left(
\frac{1}{4}\tilde{R}_{11}
-\frac{1}{48}\left(F_{b_1\dots b_4}F^{b_1\dots b_4}-F_{\alpha_1\dots \alpha_4}F^{\alpha_1\dots \alpha_4}\right)
\right).
\ee
A comparison with the Lagrangian of $D=11$ supergravity shows exact agreement modulo total derivative terms, which are also needed to transform the Chern--Simons term into a contribution to the term $F_{\alpha_1\dots \alpha_4}F^{\alpha_1\dots \alpha_4}$ that effectively flips the sign. It is important to note that all the other terms of $D=11$ supergravity cannot be expected to appear in this generalized coset model, because they are not contained in the coset $\cV\in E_{7(7)}$. After discussing fermions in the next section, we will comment on the remaining fields in section \ref{Complete2}.\footnote{For the extraction of the equations of motion, one has to keep in mind that the independent off-shell degrees of freedom are $A_{abc}$ and $A_{a_1\dots a_6}$ in this formulation.}\\

The important result of this section is that the mere quest for an $E_{7(7)}$-invariant theory with hidden $\Diff(7)$-symmetry upon dimensional reduction unambiguously leads to the dynamics of $D=11$ supergravity in the truncation to the fields and dimensions common to both theories. Note in particular that our $E_{7(7)}$-based interpretation of a part of the $D=11$ supergravity Lagrangian is different and, so to say, complementary to the one of Cremmer \& Julia \cite{CJ79}: The global $E_{7(7)}$-symmetry of $d=4$ $\cN=8$ supergravity does not act on the coordinates and hence, these four dimensions are orthogonal to the seven dimensions discussed in this section.

\end{subsection}

\end{section}

\begin{section}{Supersymmetry and fermionic dynamics}
\label{SUSY}
In section \ref{BosDyn}, we have constructed a Lagrangian from the coset degrees of freedom $\cV\in E_{7(7)}/(SU(8)/\Z_2)$. As a next step, we will address the question, whether it is possible to extend the bosonic dynamics discussed above by other fermionic fields such that there is a Gra\ss mann valued supersymmetry transformation linking the solutions of the existing bosonic theory to the fermionic dynamics, which also exhibits a hidden $\Diff(7)$ symmetry.

\begin{subsection}{Definition of the variation $\underline{\delta}$}
\label{Variation}
We start by recalling that any symmetry transformation is uniquely defined by a derivation $\delta$. Its action on the coset element $\cV$ (\ref{cosetE7}) can be decomposed in two parts in complete analogy to the Maurer--Cartan form $\cV^{-1}\cdot d\cV$ (\ref{MCf})
\be\label{SCoset0}
\cV^{-1}\cdot \delta\cV &\in &\left(\mathfrak{e}_{7(7)}\ominus \mathfrak{su}_8\right)\oplus \mathfrak{su}_8.
\ee
As in (\ref{PQTrafo}), the $\mathfrak{su}_8$-part of this coset
\be\label{SCoset2}
\Lambda &:=&\text{pr}_{\mathfrak{su}_8}\left(\cV^{-1}\delta\cV\right)
\ee
does not transform as a tensor, but as a connection. This leads to the definition of a {\sl covariant supersymmetry transformation} $\underline{\delta}$ acting on $K$-representation spaces $\psi$ in a representation $\mathbf{R}$ in complete analogy to the definition of the covariant derivative $\D$ (\ref{covder}):
\be\label{SDelta}
\underline{\delta}\psi
&:=&
\delta \psi -\mathbf{R}(\Lambda) \psi.
\ee
Thus, both $\psi$ and $\underline{\delta}\psi$ are $\mathfrak{su}_8$-tensors with respect to the $SU(8)/Z_2$-action induced by a global $E_{7(7)}$-transformation (\ref{CosTrafo2}). In contradistinction, $\delta \psi$ is not a tensor, because the compensating $k_g(\cV)\in SU(8)/\Z_2$ transformation depends on the coset field $\cV$ in general. The definition (\ref{SDelta}) also implies
\be\label{SCoset1}
\cV^{-1}\cdot \underline{\delta}\cV &\in &\mathfrak{e}_{7(7)}\ominus \mathfrak{su}_8.
\ee

As a next step, we want to specify the variation (\ref{SCoset1}). We begin by recalling that for any continuous supersymmetry transformation, there has to be a Gra\ss mann valued symmetry parameter $\eps$, in which the variation is linear \cite{Hi08}. In order to preserve covariance under the global $E_{7(7)}$ action, $\eps$ must form a representation on which at least the induced $K$-action is defined. It is obvious that the minimal dimension for a non-trivial action is $\mathbf{8}_\C$, if representations of the double cover $SU(8)$ of $K=SU(8)/\Z_2$ are included.\\

In this setting, we have to introduce fermions $\chi$ that link the variation $\left(\cV^{-1}\underline{\delta}\cV\right)^{[ABCD]}$ (\ref{SCoset1}) to the symmetry parameter $\eps^A$ with $A,B,\ldots=1,\dots,8$. By $SU(8)$ covariance, the Gra\ss mann valued fields $\chi$ must furnish the $\mathbf{56}_\C$-dimensional representation $\chi^{[ABC]}$, if derivatives of fermions are excluded. The unique symmetry transformation possible then has the form
\be\label{SCoset3}
\left(\cV^{-1}\underline{\delta}\cV\right)^{ABCD} &=:&\eps^{[A}\chi^{BCD]} + \frac{1}{4!}\e^{ABCDEFGH}\eps_{E}\chi_{FGH}
.
\ee
Adding the second term on the right hand side of (\ref{SCoset3}) is necessary to guarantee that $\cV^{-1}\underline{\delta}\cV\in \mathfrak{e}_{7(7)}\ominus \mathfrak{su}_8$ (\ref{SCoset1}) is real, in complete analogy to the equation (\ref{dual}). Again, we use the convention that changing the position of the $SU(8)$ index corresponds to a complex conjugation, e.g. $\eps^A = \eps_A^*$.\\

Furthermore note that passing to the double cover $SU(8)$ of $SU(8)/\Z_2$ does not pose any problems. The $k_g\in SU(8)/\Z_2$ action induced by $g\in E_{7(7)}$ is well-defined on any product of $SU(8)$ representations with an even number of $SU(8)$ indices. The induced $SU(8)$ action on $\eps$ and $\chi$ hence is in complete analogy to the $\Spin(d-1,1)$-actions on fermionic matter in general relativity that are induced by general coordinate transformations using the vielbein formalism.
\end{subsection}

\begin{subsection}{Fermionic $\underline{\delta}$-variations and $\Diff(7)$}
\label{SUSYDiff7}
In order to complete the definition of the variation $\underline{\delta}$, we have to fix its action on the fermion $\chi$. The requirements are linearity in the transformation parameter $\eps$ and that it should map to the degrees of freedom of the coset $\cV$ in an $SU(8)$-covariant way. Modulo nonlinear terms in either $\chi$ or derivatives, the general ansatz for this transformation is a three-parameter family with $c_0,c_1,c_2\in \R$:
\be\label{deltachi}
\underline{\delta}\chi^{ABC}\,=\,c_0\underline{\D}^{[AB}\eps^{C]}+c_1(\cP_{EF})^{EF[AB}\eps^{C]} +c_2(\cP_{EF})^{EABC}\eps^{F}.
\ee
In complete analogy to the discussion of the bosonic dynamics in section \ref{BosDyn}, the constants $c_0,c_1,c_2\in \R$ in the variation (\ref{deltachi}) will be fixed in such a way that the hidden $\Diff(7)$ symmetry is respected in a reduction to seven dimensions. In contradistinction to the variation to the bosons (\ref{SCoset3}), this requirement is non-trivial due to the appearance of derivatives in (\ref{deltachi}) that are not $\Diff(7)$-covariant in general. \\

At first, we use the decomposition of $\cP$ and $\cQ$ into $SO(7)$ representations (\ref{vGL}). Substituting these formul\ae{} into the variation $\underline{\delta}\chi$ (\ref{deltachi}), we have to look for non-trivial constants $c_0,c_1,c_2$ such that all derivatives along the $\mathbf{7}$ directions of $A_{abc}$ and of $A_{a_1\dots a_6}$ combine into the $\Diff(7)$-covariant field strengths $F$ and $\tilde{F}$ (\ref{FDefi}) respectively. This is not possible a priori.\\

However, it should be kept in mind that the explicit parametrization (\ref{cosetE73}) of the coset $\cV\in E_{7(7)}/(SU(8)/\Z_2)$ reduces the $SU(8)/Z_2$ covariance to $SO(7)$, as we have explained in section \ref{DecomGl7}. In particular, the $SO(7)$ covariance is completely sufficient for the $\Diff(7)$ covariance in the reduction $\mathbf{56}\rightarrow \mathbf{7}$. Hence, it is admissible to use the $SO(7)$ covariant intertwiners $\G^a$ (\ref{Clifford}), the $\G$-matrices, to rearrange the degrees of freedom of the $SU(8)$-representation $\chi$ into a $\Spin(7)$-representation by
\begin{subequations}\label{psichi}
\be
(\chi_a)^{C} &:=& \frac{i}{9}\left(\delta_a^b\delta^C_D +\frac{1}{8}{{{\G_a}^b}_D}^C\right){\G_b}_{AB} \chi^{ABD}\\
\Rightarrow \quad \chi^{ABC} &=& 3!i{\G^a}^{[AB}(\chi_a)^{C]}
\label{chiDefi}
\ee
\end{subequations}
with $a,b=4,\dots,10$ and $A,B,C=1,\dots,8$. It is clear that this is no truncation, because the degrees of freedom of $\chi^{ABC}$ and $(\chi_a)^{C}$ match
\beg
\binom{8}{3}\,=&56&=\,7\cdot 8.
\eeg
With this rearrangement of degrees of freedom and with the choice
\be\label{c1fix}
c_0\,\,=\,\,1, \quad c_1\,\,=\,\,-\frac{1}{2},\quad c_2\,\,=\,\,\frac23
\ee
for the constants in (\ref{deltachi}), we obtain after dropping all terms that contain partial derivatives $\frac{\p}{\p x^{mn}}$, $\frac{\p}{\p p_{mn}}$ or $\frac{\p}{\p p_{m}}$  \cite{Hi08}:
\be\label{chigl2}
\left.\Delta^{\frac14}\underline{\delta}(\chi_d)^{C}\right|_{\frac{\p}{\p x^m}}
 &=&
\p_d\left(\Delta^{-\frac{1}{4}}\eps^C\right)
+\frac{1}{4}
{\omega_{de}}^{f}
 {{{\G^{e}}_f}^C}_D
 \left(\Delta^{-\frac14}
\eps^D\right)
\nn\\
&&
+\frac{1}{144}
F_{a_1 \dots a_4}
\left(
{{{\G^{a_1\dots a_4}}_d}^{C}}_D 
   -8\delta_d^{a_1}{{\G^{a_2\dots a_4}}^{C}}_D
   \right) \left(\Delta^{-\frac14}\eps^D\right)
\nn\\
&&-\frac{i}{7!6}\dF
{{\G_{d}}^{C}}_D
 \left(\Delta^{-\frac14}\eps^D\right).
\ee
In this formula, we used the abbreviations $\omega$ (\ref{Abk}), $\Delta$ (\ref{sigmaDefi2}), $F$ and $\dF$ (\ref{FDefi}). To obtain a $\Diff(7)$ covariant transformation, the fields $\eps$ and $\chi$ are rescaled by the determinant factors $\Delta$ of ${e_m}^a$ in analogy to a Weyl rescaling (\ref{Weyl2}). This is possible due to the following equality modulo quadratic terms in $\chi$ that were neglected in the definition of the variation $\underline{\delta}\chi$ (\ref{deltachi}) anyway:
\be\label{equiv4}
\Delta^{\frac14}\underline{\delta}(\chi_d)^{C} &=& \underline{\delta}\left(
\Delta^{\frac14}(\chi_d)^{C}\right) +\mathcal{O}(\chi^2)
\ee
The statement (\ref{equiv4}) follows from the fact that $\Delta$ is part of the coset $\cV$ (\ref{cosetE73}) and hence its variation under $\underline{\delta}$ can be deduced from (\ref{SCoset3}). We will provide explicit formul\ae{} for the rescaling of $\chi$ and $\eps$ in section \ref{CS2}.
\end{subsection}

\begin{subsection}{$\underline{\delta}$ and the additional four dimensions}
\label{SUSYCos}
In the end, we should be interested in a complete picture of the generalized coset dynamics. In particular, we want to discuss a theory that contains the coset $\cV\in E_{7(7)}/(SU(8)/\Z_2)$-coset, but whose reduction is $\Diff(7)$ invariant by itself. The result of section \ref{restdiff7} was that this requirement naturally leads to four additional dimensions with the additional fields ${e_\mu}^\alpha$ and ${B_\mu}^a$ (\ref{Weyl2}). The minimal $E_{7(7)}$-covariant extension that allows to include ${e_\mu}^\alpha$ and ${B_\mu}^a$ in a coset description is the group presented in section \ref{GE7}
\be\label{Ggesamt0}
G&=&\big(Gl(4)\times E_{7(7)}\big)\ltimes \cN_{(4,\,56)}.
\ee
This ansatz is promising, because it allows to construct a coset with the de Wit--Nicolai covariance group $SO(3,1)\times SU(8)/Z_2$ of $D=11$ supergravity:
\be\label{Ggesamt}
\cV&\in& G/\big(SO(3,1)\times SU(8)/Z_2\big).
\ee
The additional degrees of freedom of the four dimensional vielbein ${e_\mu}^\alpha$ (\ref{Weyl2}) parametrize the top left block of the matrix representation of $\cV$ in (\ref{GenV}) and the field ${B_\mu}^a$ (\ref{Weyl2}) is contained in the top right block of (\ref{GenV}). The complete parametrization of the coset $\cV$ (\ref{Ggesamt}) in terms of $Gl(7)$ representations will be provided in section \ref{Complete2}.\\

The definition of the variation $\underline{\delta}$ from section \ref{Variation} has to be adapted to this extended setting. At first, the relation (\ref{SCoset1}) is replaced by
\be\label{SCoset2b}
\cV^{-1}\underline{\delta}\cV &\in &\left(\mathfrak{e}_{7(7)}\ominus \mathfrak{su}_8\right)
\oplus \left(\mathfrak{gl}_4\ominus \mathfrak{so}_{(3,1)}\right)\oplus  \mathfrak{n}_{(4,56)}.
\ee

As before, we want to assume a non-trivial realization of the covariance group on the continuous symmetry parameter $\eps$. Then, the lowest real dimension is $\mathbf{32}$ due to the following reason: Passing to the covering $\Spin(3,1)\times SU(8)$ of the de Wit--Nicolai group, a $\Spin(3,1)$-representation is constructed from the Clifford algebra (\ref{Clifford})
\be\label{Clifford2}
\{\g^\alpha,\g^\beta\} &=& \eta^{\alpha\beta}
\ee
with the Minkowksi metric $\eta=\diag(-1,1,1,1)$ that has a representation as real matrices $\g^\alpha\in \R^{4\times 4}$. It is a standard observation that the real matrix
\be\label{gamma5}
\g_5&:=&\g^0\g^1\g^2\g^3
\ee
squares to $-\id_4$.\footnote{With the definition $\e^{0\,1\,2\,3}=1$, the definition (\ref{gamma5}) implies $\g_5\e^{\alpha_1\dots \alpha_4}=\g^{\alpha_1\dots \alpha_4}$.} This is the reason why the vector space
\be\label{48}
\mathbf{4}_\R\otimes \mathbf{8}_\R
\ee
forms a representation space of $\Spin(3,1)\times SU(8)$, with $\Spin(3,1)$ acting irreducibly on the first factor $\mathbf{4}_\R$ and the so-called ``chiral'' $SU(8)$ \cite{CJ79} acting on both factors in (\ref{48}) making use of $\g_5$ as the imaginary unit
\be\label{gamma5Act}
\g_5\eps &=& i\eps.
\ee

To put it in other words, this identification of $\g_5$ with the imaginary unit $i$ provides an embedding of $\C$ in $\R^4$. Following the lines of section \ref{Variation}, $\eps\in \mathbf{4}_\R\otimes \mathbf{8}_\R$ implies that the $\mathbf{56}_\C$-dimensional representation of the fermions $\chi$ now has to be extended to the real representation $\mathbf{4}_\R\otimes \mathbf{56}_\R$ of $\Spin(3,1)\times SU(8)$, on which $SU(8)$ acts in a chiral way with (\ref{gamma5Act}) again.\\

Next, we introduce the Majorana conjugate of $\eps\in \mathbf{4}_\R\otimes \mathbf{8}_\R$ in order to keep the notation as simple as possible:
\be\label{Majorana}
\bar{\eps}^A &:=& (\eps^t)^A\g^0.
\ee
This allows to suppress the spinor indices of $\mathbf{4}_\R$, e.g. the matrix indices of $\g^\alpha$, in the following. It should be noted that the position of the $SU(8)$ index is not affected by the four-dimensional transposition $t$.\\

It is obvious from equation (\ref{SCoset2b}) that the definition (\ref{SCoset3}) is to be understood as a projection of $\cV^{-1}\underline{\delta}\cV$ (\ref{SCoset2b}) on $\mathfrak{e}_{7(7)}\ominus \mathfrak{su}_{8}$. Due to the extension of $\eps\in \mathbf{8}_\C$ to $\eps\in\mathbf{4}_\R\otimes \mathbf{8}_\R$ (\ref{48}), the equation (\ref{SCoset3}) now takes the form
\be\label{SCoset4}
\left(\cV^{-1}\underline{\delta}\cV\right)^{ABCD} &=&\bar{\eps}^{[A}\chi^{BCD]} + \frac{1}{4!}\e^{ABCDEFGH}\bar{\eps}_{A}\chi_{BCD}.
\ee
Raising or lowering the $SU(8)$ indices of the fermions is hence equivalent to replacing $\g_5$ by $-\g_5$ in accordance with (\ref{gamma5Act}).\\

Before comparing these formul\ae{} to supergravity, we extract the variations of the $Gl(7)$ representations ${e_m}^a$, $A_{abc}$ and $A_{a_1\dots a_6}$ (\ref{cosetE73}) from the one of the coset $\cV$ (\ref{SCoset4}). A short calculation \cite{Hi08} leads to
\begin{subequations}\label{psiDefi}
\be
{e_b}^m\delta {\left.e\right._m}^c&=& 
i\bar{\eps}^C{\G^c}_{CD}(\chi_b)^D + \text{c.c.}\\
{e_{a_1}}^{m_1}\cdots{e_{a_3}}^{m_3}
\delta A_{m_1\dots m_3}
&=&
-\frac{3i}{2}\bar{\eps}^C{\G_{[a_1a_2}}_{CD}(\chi_{a_3]})^D + \text{c.c.}\\
{e_{a_1}}^{m_1}\cdots{e_{a_6}}^{m_6}
\delta A_{m_1\dots m_6}
&=&
  -3i\bar{\eps}^C{\G_{[a_1\dots a_5}}_{CD}(\chi_{a_6]})^D + \text{c.c.}
 \\
 &&+20{e_{a_1}}^{m_1}\cdots{e_{a_6}}^{m_6}
A_{[m_1\dots m_3}\delta A_{m_4\dots m_6]}\nn
 \ee
\end{subequations}
with the definitions for $i=3,6$
\be\label{vSym7}
A_{m_1\dots m_i} &:=& {e_{m_1}}^{a_1}\cdots{e_{m_i}}^{a_i}A_{a_1\dots a_i}.
\ee
\end{subsection}

\begin{subsection}{Second comparison to $\mathbf{D=11}$ supergravity}
\label{CS2}
After having established a connection between the coset degrees of freedom of $\cV\in E_{7(7)}/(SU(8)/\Z_2)$ and the bosonic fields of supergravity in section \ref{CS1}, it is natural to identify the variation $\delta$ (\ref{SDelta}) with the supersymmetry variation $\delta_{\ep}$ (\ref{Trafo1neu}) of supergravity. Its $32$ real dimensional transformation parameter $\ep$ is linked to the transformation parameter $\eps\in \mathbf{4}_\R\otimes \mathbf{56}_\R$ following Cremmer \& Julia \cite{CJ79}:
\begin{subequations}\label{rescale4}
\be
{\eps}^C &=& \frac{1}{2}\sqrt{-\g_5}\Delta^{+\frac{1}{4}}\left(\id_4 -i\g_5\right){\ep}^C \\
(\chi_a)^C&=&\frac{1}{2}\sqrt{-\g_5}\Delta^{-\frac{1}{4}}\left(\id_4 -i\g_5\right)(\psi_{a})^C
\label{rescale4b}\\
\text{with}\quad \sqrt{-\g_5}&:=& \frac{1}{\sqrt{2}}\left(\id_4 -\g_5\right).
\nn
\ee
\end{subequations}
Furthermore, the fermion $\chi\in \mathbf{56}_\R\otimes \mathbf{4}_\R$ is identified with the gravitino $\psi$ of supergravity for the vector indices $a=4,\dots,10$ by (\ref{rescale4b}). It is nice to observe that the same rescaling with the determinant $\Delta$ (\ref{sigmaDefi2}) used by Cremmer \& Julia \cite{CJ79} also is the correct choice in order to obtain $\Diff(7)$-covariance in the variation $\underline{\delta}\chi$ (\ref{chigl2}).\footnote{We emphasize again that the present discussion of the seven dimensional reduction of $D=11$ supergravity is complementary to the four dimensional one of Cremmer \& Julia discussed in \cite{CJ79}. Relations without derivatives such as (\ref{SCoset3}) can also be found in their article, of course, in contrast to most of the terms in e.g. the variation $\underline{\delta}\chi$ (\ref{chigl2}), however.}\\

The identification (\ref{rescale4}) finally allows to compare the variations of the coset $\cV$ (\ref{psiDefi},\,\ref{SCoset4}) and of $\chi$ (\ref{chigl2}) to the ones of $D=11$ supergravity (\ref{Trafo1neu}). In the truncation defined by (\ref{Weyl2},\,\ref{assump}), these exactly match keeping in mind the standard decomposition of $\tilde{\G}$-matrices in eleven dimensions \cite{Hi08}
\begin{subequations}\label{gammadec}
\be
\tilde{\G}_\alpha &=& \g_\alpha\otimes \id_{8}\quad \,\text{for }\alpha\,=\,0,\dots,3,\label{G4Defi}\\
\tilde{\G}_a &=& \frac{\g_5}{i}\otimes \G_a\quad \text{for }a\,=\,4,\dots,10.
\label{G5Defi}
\ee
\end{subequations}
Note in particular that the normalizations of the coset fields $A_{abc}$ and $A_{a_1\dots a_6}$ are fixed by the comparison of the bosonic actions in section \ref{CS1}. Then it is a non-trivial result that the numerical constants in the supersymmetry variations of the fermions (\ref{chigl2}) and of the bosons (\ref{psiDefi}) exactly agree with the ones of $D=11$ supergravity (\ref{Trafo1neu}) in the present truncation \cite{Hi08}. \\

Together with the group theoretical argument from section \ref{restdiff7}, this is a strong indication that the generalized coset dynamics of $E_{7(7)}$ is related to $D=11$ supergravity.\\

There is one additional fermion in supergravity whose counterpart in the generalized coset dynamics has not been discussed so far, the gravitino $\psi_\alpha$ with the vector index $\alpha=0,\dots,3$. Splitting the $\mathbf{32}$ dimensional spinor representation space of $\psi_\alpha$ into the product $\mathbf{4}_\R\times \mathbf{8}_\R$ as in section \ref{SUSYCos}, it is natural to use the same definition as Cremmer \& Julia \cite{CJ79}
\be\label{rescale5}
(\chi_\alpha)^C\,:=\,\frac{1}{2}\sqrt{-\g_5}\Delta^{-\frac{1}{4}}\left(\id_4 -i\g_5\right)\left((\psi_\alpha)^C +\frac{i}{2}\g_5\g_\alpha {{\G^a}^C}_D(\psi_a)^D\right).
\ee
The matrix $\g_5$ obviously serves as a complex structure $\gamma_5 (\chi_\alpha)^C = i (\chi_\alpha)^C$ as in (\ref{gamma5Act}). Hence, $(\chi_\alpha)^C$ forms a representation space of $\Spin(3,1)\times SU(8)$. It remains to check whether the $SU(8)$-covariant variation of this additional field allows for a hidden symmetry $\Diff(7)$ upon a Kaluza--Klein reduction $\mathbf{56}\rightarrow \mathbf{7}$. This will be proved next.
\end{subsection}

\begin{subsection}{Variation of $\chi_\alpha$}
\label{MoreSUSYSU}
The procedure to obtain the symmetry transformation $\underline{\delta}$ of $(\chi_\alpha)^C$ is completely analogous to the one used for the field $\chi^{ABC}$ in (\ref{deltachi}). By $SU(8)$-covariance and neglecting derivatives along the additional four directions, the general Ansatz (modulo non-linear terms in derivatives or $\chi$) is with $e_1,e_2\in \R$
\be\label{deltachi3}
\underline{\delta} (\chi_\alpha)_A 
&=&
 \g_\alpha\left(e_1\underline{\D}_{AB}\eps^B +e_2\big(\cP^{CD}\big)_{ABCD}\eps^B\right).
\ee
This equation exhibits $\Diff(7)$-covariance in the reduction $\mathbf{56}\rightarrow \mathbf{7}$ for the constants
\beg
e_2&=& -\frac{1}{2}e_1.
\eeg
Setting $e_1=\frac{1}{12}$ fixes the normalization of $(\chi_\alpha)^C$ in a suitable way for a comparison to $D=11$ supergravity. In a final step to simplify the notation, we suppress the $SU(8)$ indices of $\chi$, $\eps$ and of the matrices $\G$, if the way to contract them is unambiguous. Adding a star to fermions with lowered $SU(8)$ indices in order to distinguish them from the ones with raised indices, we obtain for (\ref{deltachi3}) with the formul\ae{} for $\cP$ and $\cQ$ (\ref{vGL}) in the same truncation as in (\ref{chigl2}):
\beg
\left.\underline{\delta} \chi^*_\alpha \right|_{\frac{\p}{\p x^m}}
&=&
\frac{i}{2}\g_\alpha
\Delta^{-\frac{1}{2}}\left[
\G^a\left\{\p_a\eps
+\frac{1}{4}
{\omega_{ab}}^{c}
 {\G^{b}}_c
 \eps
-\frac{3}{4}e^{-1}\p_a e \eps
\right\}
\right.\\
&&\left.
-\frac{1}{48}F_{a_1\dots a_4}
\G^{a_1\dots a_4}
\eps
  -\frac{i}{7!2}\dF\eps  \right] 
\eeg
This variation exactly agrees with the supersymmetry transformation of the remaining gravitino degrees of freedom of $D=11$ supergravity (\ref{Trafo1neu}) in the present truncation.
\end{subsection}

\begin{subsection}{Fermionic Dynamics}
\label{SummFerm}
After having defined an $SU(8)$-covariant variation that links the degrees of freedom of the coset $\cV$ to the fermions $\chi$ (\ref{deltachi},\,\ref{SCoset4},\,\ref{deltachi3}), we will investigate whether their dynamics can be defined in such a way that a hidden $\Diff(7)$ symmetry appears in the Kaluza--Klein reduction $\mathbf{56}\rightarrow \mathbf{7}$ again.\\

We will follow the same pattern used for the action $S$ of the coset degrees of freedom (\ref{Action7}) in section \ref{Lagr}: At first, we construct the general $SU(8)$-covariant Lagrangian in $56$ dimensions that is linear in derivatives and at most quadratic in $\chi$, where derivatives along the additional four directions are neglected as before. Then, we fix the constants by requiring the hidden $\Diff(7)$ symmetry to appear in the Kaluza--Klein reduction $\mathbf{56}\rightarrow \mathbf{7}$, taking into account the Weyl rescaling of the additional four dimensions (\ref{Weyl2}). Thus, one is led to the following Lagrangian:\footnote{It is clear that only the constants inside the square brackets are fixed by requiring a hidden $\Diff(7)$ covariance. The numerical factors linking the square brackets to each other are fixed by a comparison to $D=11$ supergravity. We will comment on a possible group theoretic origin of these constants at the end of section \ref{CONCL}.}
\beg
\mathcal{L}_{\text{fermions}}
 	 &=&
-\frac{1}{12}\bar{\chi}^{ABC}\left[
\underline{\D}^{DE}\chi^{FGH}
-\frac32\left(\cP_{JK}\right)^{DEJK}\chi^{FGH}
\right.\\
&&\left.
+2\left(\cP_{JK}\right)^{DEJF}\chi^{KGH}
\right]
\e_{ABCDEFGH}\\
&&+\frac{1}{96}\bar{\chi}_{ABC}\g^\alpha
\left[
 \underline{\D}^{AB}(\chi_\alpha)^C
-\frac12\left(\cP_{JK}\right)^{ABJK}(\chi_\alpha)^C
\right.\\
&&\left.
+\frac23\left(\cP_{JK}\right)^{JABC}(\chi_\alpha)^K
 \right]
\\
&&+\frac{1}{12}(\bar{\chi}_\alpha)^{A}\g^{\alpha\beta}
\left[
 \underline{\D}_{AB}(\chi_\beta)^B
 -\frac12\left(\cP^{JK}\right)_{ABJK}(\chi_\beta)^B
 \right]\\
&&+\frac{1}{96}(\bar{\chi}_\alpha)_{A}\g^{\alpha} 
\left[
\underline{\D}_{BC}\chi^{ABC}
 -\frac{11}{6}\left(\cP^{JK}\right)_{JKBC}\chi^{ABC}
\right.\\
&&\left.
-2\left(\cP^{J[A}\right)_{JKBC}\chi^{BC]K}
\right]
 +\text{c.c.}
	 \eeg
	 
To prove the hidden $\Diff(7)$-symmetry, we perform the Kaluza--Klein reduction $\mathbf{56}\rightarrow \mathbf{7}$ (\ref{egl7}) of this Lagrangian. After substituting the $\mathfrak{so}_7$ decompositions of the bosonic fields (\ref{vGL}) and of the fermions (\ref{psichi}) we obtain without dropping total derivative terms nor making use of the anticommutativity of $\chi$:

\beg
&&\left.\mathcal{L}_{\text{fermions}}\right|_{\text{KK}}\\
&=&\Delta^{-\frac{1}{2}}\left[
\bar{\chi}_k\left(\frac{3}{2i}{\G^{(j}}\eta^{dk)} 
\right)
\left\{
\p_d \chi_j
+\frac{1}{4}{\omega_{de}}^{f} {\G^e}_f \chi_j
 +{\omega_{dj}}^f \chi_f
 -\frac{3}{4}
\chi_f\Delta^{-1}\p_d\Delta 
  \right\} 
  \right.\\
&&+\frac{1}{72}F_{b_1\dots b_4}
\bar{\chi}_k
 \left\{
{\e^{b_2\dots b_4(j}}_{rst}\eta^{k)b_1}
-\frac{1}{8}{\e^{b_1\dots b_4}}_{rst}\eta^{jk}
 \right\}{\G^{rst}}
\chi_j \\
&&\left.
-\frac{1}{2i}F^{b_1\dots b_4}
\bar{\chi}_{b_4}
{\G_{b_1b_2}}
\chi_{b_3}
+\frac{1}{7!}\dF
\bar{\chi}_k\left(
{\G^{jk}}
 -\frac{3}{4}
\eta^{jk}
  \right)
 \chi_j
 \right]\\
&&
+\Delta^{-\frac{1}{2}}
 \bar{\chi}^*_k \g^{\gamma}
 \left[
 \left(\frac{3}{4}
  \eta^{kc}
+\frac{1}{4}\G^{kc}
 \right)
   \left\{
\p_c \chi_\gamma +\frac{1}{4}{\omega_{ce}}^{f}{\G^{e}}_f\chi_\gamma
 -\frac{1}{4}\chi_\gamma \Delta^{-1}\p_c \Delta 
 \right\}
   \right.\\
	 &&\left.-\frac{1}{64}\left(
	 {\G^k}\G^{b_1\dots b_4}
	 -\frac{2}{3}\G^{b_1\dots b_4k}
	 \right)\chi_{\gamma}F_{b_1\dots b_4} 
	 -\frac{3i}{7!8}
		 \G^k\chi_{\gamma}\dF\right]\\
	 &&+\Delta^{-\frac{1}{2}}
 \bar{\chi}_\beta\g^{\beta\gamma}\left[ \frac{i}{2}\G^c 
 \left\{
\p_c \chi_\gamma +\frac{1}{4}{\omega_{ce}}^{f}{\G^{e}}_f\chi_\gamma
 -\frac{3}{4}\chi_\gamma \Delta^{-1}\p_c \Delta 
 \right\} \right.\\
	 &&\left.
	 -\frac{i}{96} \G^{b_1\dots b_4}\chi_{\gamma}
	 F_{b_1\dots b_4}
	 +\frac{1}{7!4}\chi_{\gamma}
	 \dF \right]\\
 &&
+\Delta^{-\frac{1}{2}}\bar{\chi}^*_\beta \g^{\beta}
\left[
\left(\frac{3}{4}
  \eta^{cj}
+\frac{1}{4}\G^{cj}
 \right)
 \left\{
\p_c \chi_j +\frac{1}{4}{\omega_{ce}}^{f}{\G^{e}}_f\chi_j
 +{\omega_{cj}}^f\chi_f
 \right.\right.\\
	 &&\left.\left.
 -\frac{5}{4}\chi_j \Delta^{-1}\p_c \Delta 
 \right\}
	 +\frac{1}{64}\left(
	 \G^{b_1\dots b_4}\G^j
	 -\frac{2}{3}\G^{b_1\dots b_4j}
	 	 \right)\chi_jF_{b_1\dots b_4} 
	 -\frac{3i}{7!8}\G^j\chi_j \dF
\right]\\
&&
 +\text{c.c.}
	 \eeg
	 
In order to simplify the notation, the $SU(8)$ indices of $\chi$ and $\G$ have been dropped, which is possible with the star notation introduced in section \ref{MoreSUSYSU}. The important result is that $\mathcal{L}_{\text{fermions}}|_{\text{KK}}$ exactly coincides with the Kaluza--Klein reduction of the fermionic part of the $D=11$ supergravity Lagrangian (\ref{Actionneu}) from section \ref{SUGRA}
\beg
\mathcal{L}_{\text{fermions}}
 &=&
 \det(E)\left(-\frac{1}{2}\bar{\psi}_{\ca_1}\tilde{\G}^{\ca_1\dots \ca_3}\nabla_{\ca_2}\psi_{\ca_3} 
	\right.\\
	&&\left.
	 -\frac{1}{96}\left( \bar{\psi}_{\ca_5}\tilde{\G}^{\ca_1\dots \ca_6}\psi_{\ca_6} +12\bar{\psi}^{\ca_1}\tilde{\G}^{\ca_2\ca_3}\psi^{\ca_4}\right)F_{\ca_1\dots \ca_4} 
	 \right),
\eeg
	 if we firstly split the summations $\ca=0,\ldots,10$ into $\alpha=0,\ldots,3$ and $a=4,\ldots,10$, secondly use the decomposition of $\tilde{\G}$-matrices (\ref{gammadec}), thirdly substitute $\chi$ for $\psi$ (\ref{rescale4},\,\ref{rescale5}), fourthly drop all derivatives $\frac{\p}{\p x^\alpha}$ with $\alpha=0,\dots,3$ and finally use the simplifying assumptions (\ref{assump}) for the vielbein $E$ in eleven dimensions (\ref{Weyl2}). This completes the proof that a hidden symmetry $\Diff(7)$ also appears in the fermionic dynamics, if the Weyl rescaling of the additional four dimensions (\ref{Weyl2}) is taken into account.
\end{subsection}
\end{section}

\begin{section}{Exceptional geometry}
\label{Complete}
The hidden $\Diff(7)$-symmetry of the bosonic and the fermionic dynamics in sections \ref{restdiff7} and \ref{SummFerm} respectively provided evidence that adding four dimensions to the $56$ dimensional generalized coset dynamics may result in an interesting structure. In this section, the generalized coset picture will be completed with the bosonic fields that are naturally linked to the additional four dimensions following the discussion of section \ref{SUSYCos}. We will conclude with the geometric interpretation and comments on the literature.

\begin{subsection}{A glance at the complete theory}
\label{Complete2}
Up to now, we have only motivated the Minkowskian signature for the additional four dimensions by a comparison with $D=11$ supergravity. However, this also is preferred from a pure group theoretical point of view, if we require the dimension of the supersymmetry parameter $\eps$ to be minimal.\\

This is due to the fact that for the other two independent signatures $(\!++++\!)$ and $(\!++--\!)$, there either is no Majorana representation of the Clifford matrices $\g_\alpha$ in $\R^{4\times 4}$, or $\g_5$ (\ref{gamma5}) would square to $+\id_4$ implying that the real matrix $\g_5$ would not provide a complex structure (\ref{gamma5Act}). Therefore, the minimal real dimension of the non-trivial representation space of both $Spin(4)\times SU(8)$ and $Spin(2,2)\times SU(8)$ would be $\mathbf{4}_\R\times 2\cdot \mathbf{8}_\R>\mathbf{32}_\R$, which would be in contradiction to the maximality of $d=4$ $\cN=8$ supergravity. The Minkowskian choice for the signature then leads to the coset (\ref{Ggesamt})
\be\label{Ggesamt2}
\cV&\in& G/\big(SO(3,1)\times SU(8)/Z_2\big)
\ee
of the group $G=\big(Gl(4)\times E_{7(7)}\big)\ltimes \cN_{(4,\,56)}$ (\ref{Ggesamt0}), whose matrix representation as $60\times 60$ matrices is (\ref{GenV}):
\beg
		\left(\begin{tabular}{c|c}
	$Gl(4)
	$ & $*_{\text{\tiny4x56}}$\\
	\hline
	$0$												& $E_{7(7)}
	$
	\end{tabular}
	\right)
	\eeg
	
		As explained in section \ref{SUSYCos}, the vierbein ${e_\mu}^\alpha$ parametrizes the coset $Gl(4)/SO(3,1)$. The $Gl(4)\times Gl(7)$-decomposition of the $\mathbf{4}_\R\times \mathbf{56}_\R$ additional off-shell degrees of freedom in the coset $\cV$ (\ref{Ggesamt2}) suggests an identification with the following supergravity fields:
	\begin{enumerate}
	\item $4\times 7$-part ${B_\mu}^a$ of the vielbein $E$ (\ref{Vielbdecom}),
	\item $4\times 21$-part $A_{\mu ab}$ of the three-form potential $A$ (\ref{F4a}),
	\item $4\times 21$-part $\tilde{A}_{\mu a_1\dots a_5}$ of the dual six-form potential $\tilde{A}$ (\ref{7Form})
	\item and an additional field $C_\mu{}^a$.
\end{enumerate}
\newpage
The duality relation (\ref{4Form}) guarantees that the three- and the six-form potentials $A_{\mu ab}$ and  $\tilde{A}_{\mu a_1\dots a_5}$ are independent variables in a Lagrangian, but what is the counterpart of ${C_\mu}^a$ in $D=11$ supergravity?\\

In a Kaluza--Klein reduction of the complete generalized coset dynamics in sixty dimensions down to eleven dimensions, the $4\times 7$ degrees of freedom of ${C_\mu}^a$ apparently rule out an off-shell $\Diff(11)$-covariance. It is however not unlikely that $\Diff(11)$ is an on-shell symmetry of the reduced theory. This possibility is backed up by the similar properties of the generalized coset dynamics and $D=11$ supergravity: Apart from their identical dynamics and supersymmetry variations in the seven dimensional sector, which was discussed in this article, both theories exhibit a global on-shell $E_{7(7)}$-invariance upon a Kaluza--Klein reduction to four dimensions. The unique way to settle this issue appears to be a comparison of the complete generalized coset dynamics to $D=11$ supergravity, which is beyond the scope of the present article, however.
\end{subsection}

\begin{subsection}{Geometric interpretation}
\label{Geom}
At the end of section \ref{KK}, we explained that an ``exceptional geometry'' would be necessary to consistently define the theory on more general spaces than the vector space $\R^{56}\oplus \R^4$ of section \ref{GE7}. A first example of such a constrained geometry was provided by symplectic geometry in section \ref{Sp}: It is consistent (though not general) to assume that the vielbein $\cV$ on a symplectic manifold $(\cM^{2n},\Omega)$ is parametrized by the degrees of freedom of the coset $Sp(2n)/U(n)$.\\

Let us discuss this example in more detail. It is well known that any symplectic manifold $(\cM^{2n},\Omega)$ has a Lagrangian submanifold $\cM^n_L$ of dimension $n$ \cite{McDS95}. Furthermore, the cotangent bundle over the Lagrangian submanifold $T^*\cM^n_L$ is diffeomorphic to an open neighbourhood of the Lagrangian submanifold $\cM^n_L$ in $(\cM^{2n},\Omega)$. Therefore, it is clear that from a local point of view, the vielbein $\cV\in Sp(2n)/U(n)$ or the corresponding metric $g$ (\ref{Metrik}) can equivalently be defined on the $2n$-dimensional cotangent bundle $T^*\cM^n_L$, i.e. for every $x\in\cM^n_L$ and $v\in T_x^*\cM^n_L$, $g$ is the mapping
\be\label{gDefi4}
g(x,v):T_{(x,v)}\left(T_x^*\cM^n_L\right) \otimes T_{(x,v)}\left(T_x^*\cM^n_L\right)&\longrightarrow & \R.
\ee

This is the setting that we want to generalize. Instead of the cotangent bundle $T^*\cM^n_L$ over the Lagrangian submanifold, consider a vector bundle $E$ over a four dimensional manifold $\cM^4$ with $56$ dimensional fibre $E_x$ for any $x\in \cM^4$ and with the canonical projection
\be\label{pi}
\pi : E&\longrightarrow & \cM^4.
\ee

Furthermore, endow this vector bundle $E$ with an $E_{7(7)}$-structure. For every $x\in \cM^4$, this effectively reduces the group $Gl(56)$ of endomorphisms
\beg
\vp_x:E_x & \longrightarrow & E_x
\eeg
to $E_{7(7)}$ \cite{J00}.\footnote{An equivalent way to restrict the morphisms $\vp:E\rightarrow E$ is to require that the maps $\vp_x: E_x\rightarrow E_x$ preserve a symplectic form $\Omega$ and the quartic symmetric tensor $Q$ which is the invariant tensor of $E_{7(7)}$ \cite{Hi08}.} Next, we define a metric for every $x\in \cM^4$ and $v\in E_x$ by the non-degenerate symmetric mapping
\be\label{gDefi5}
g(x,v):T_{(x,v)}E \otimes T_{(x,v)}E&\longrightarrow & \R.
\ee

As for the symplectic case, we want to make use of the geometric structure to consistently reduce the off-shell degrees of freedom of the associated vielbein $\cV$ (\ref{VielbeinG},\,\ref{Metrik}). Due to the $E_{7(7)}$-structure of $E$, the vector bundle morphisms $\vp:E\rightarrow E$ do not violate the following restriction on the vielbein $\cV$, presented in its representation $\mathbf{R}$ as a $60\times 60$ matrix (\ref{GenV}):
\beg
	\mathbf{R}(\cV)
	&\in&
	\left(\begin{tabular}{c|c}
	$Gl(4)
	$ & $*_{\text{\tiny4x56}}$\\
	\hline
	$0$												& $E_{7(7)}$
	\end{tabular}
	\right).
	\eeg
	
This exactly is the ansatz used for the generalized coset dynamics in the sections \ref{GE7}, \ref{SUSYCos} and \ref{Complete2}. Finally, the signature of the metric $g$ (\ref{gDefi5}) is fixed to be Minkowksian, as well as the one of its canonical restriction to $T_{x}\cM^4 \otimes T_{x}\cM^4$. This completes the geometric setting for the generalized coset dynamics of section \ref{BosDyn}.\\

The fermions $\chi$ and $\eps$ can also be encoded in the geometrical picture. They parametrize sections of vector bundles over the sixty dimensional manifold $E$ (\ref{pi}). A general vector bundle morphism $\vp:E \rightarrow E$ respecting the $E_{7(7)}$ structure induces a $\Spin(3,1)\times SU(8)$ action on $\chi$ and $\eps$ in complete analogy to the standard $\Spin(3,1)$-action on the spin bundle induced by a coordinate transformation $\Diff(4)$ in general relativity \cite{Hi08}.
\end{subsection}

\begin{subsection}{Exceptional geometry and comments on the literature}
\label{Literature}
The idea to add more dimensions to $D=11$ supergravity has been discussed before. To our knowledge, the number $60$ appeared for the first time in de Wit \& Nicolai's review as a conjecture for a ``BPS-extended supergravity'' \cite{dWN00}. In analogy to the discussion of the $E_{8(8)}$-case \cite{KNS00}, they baptized the underlying hypothetical geometrical structure {\textit{exceptional geometry}}.\\

Since the generalized coset dynamics in sixty dimensions with hidden $\Diff(7)$ symmetry perfectly agrees with $D=11$ supergravity for the comparable fields so far, we will henceforth adopt this name and define the {\textit{dynamics of exceptional geometry}} to be described by the extensions of the Lagrangians of the sections \ref{Lagr} and \ref{SummFerm} to the entire sixty dimensional setting.\\

Before concluding, we would like to emphasize the difference of the present exceptional geometry to Hull's definition of an ``M-geometry on a seven dimensional manifold $\mathcal{H}$'' \cite{Hu07} or Pacheco \& Waldram's ``exceptional generalized geometry'' (EGG) \cite{PW08} \`a la Hitchin. All settings contain a vector bundle with structure group $E_{7(7)}$, but the base manifold is of different dimension. In particular, ``M-geometries'', EGGs and ``U-folds'' by definition \cite{Hu07,PW08} possess a manifest diffeomorphism symmetry $\Diff(7)\subset \Diff(11)$ in contradistinction to the exceptional geometry in sixty dimensions, whose hidden symmetry $\Diff(7)$ only appears in a truncation to eleven dimensions. However, it would be interesting to check whether these constructions are related.
\end{subsection}
\end{section}

\begin{section}{Conclusion and outlook}
\label{CONCL}
The logic of this paper has been the following:
\begin{enumerate}
	\item In section \ref{BosDyn}, we have applied the generalized coset dynamics of section \ref{CosDyn} to the Lie group $G=E_{7(7)}$. It turns out that there is a Lagrangian depending on $56$ dimensions whose Kaluza--Klein reduction to seven dimensions can be made $\Diff(7)$-covariant, if and only if $d=4$ additional dimensions are coupled to the system in the sense explained in section \ref{restdiff7}. A first comparison with the bosonic part of the Lagrangian of $D=11$ supergravity in section \ref{CS1} shows perfect agreement for the fields under consideration.
\item The section \ref{SUSY} discussed the possibility of an $E_{7(7)}$-covariant variation $\underline{\delta}$ on the coset degrees of freedom $\cV\in E_{7(7)}/(SU(8)/\Z_2)$ which results in the definition of the fermions $\chi^{ABC}$ (\ref{SCoset3}). To lowest order in derivatives and fermions, $\underline{\delta}\chi$ was uniquely fixed by requiring the hidden $\Diff(7)$-covariance of the bosonic part to persist. The same is true for the fermionic dynamics, if the Weyl rescaling of the additional four dimensions is taken into account again. A second comparison with the supersymmetry variation of $D=11$ supergravity and its fermionic Lagrangian also shows perfect agreement, even if the degrees of freedom of the gravitino $\psi_\alpha$ with $\alpha=0,\dots,3$ are included.
\item Section \ref{Complete} finally provided a glance at the complete theory in sixty dimensions and its geometrical interpretation in terms of an exceptional geometry.
\end{enumerate}
The agreement with $D=11$ supergravity in the compared sector together with the manifest $E_{7(7)}$-invariance of sixty dimensional exceptional geometry leads to the suspicion that its equations of motion could be preserved under the $\mathbf{32}_\R$ dimensional supersymmetry variation $\underline{\delta}$ of section \ref{SUSY}. Hence, its truncation to four dimensions would immediately provide a Lagrangian formulation of $\cN=8$ $d=4$ supergravity with manifest, off-shell $E_{7(7)}$-invariance. This would also be a strong argument in favour of a hidden symmmetry $\Diff(11)$ of the sixty dimensional exceptional geometry in a truncation to eleven dimensions as discussed in section \ref{Complete2}.\\

Following the same line of argumentation, it would then also be likely that $\Diff(10)\times Sl(2)$ is a hidden symmetry of exceptional geometry and that the dynamics of $IIB$ supergravity are contained in the ones of exceptional geometry, too. This possibility is linked to the observation that not all $Gl$ subgroups of $E_{7(7)}$ are contained in its $Gl(7)$ subgroup. This is in particular the case for $Gl(6)\times Sl(2)$. \\

We want to emphasize that the closure of the supersymmetry algebra has not been used for the construction of the dynamics in this article. Nevertheless, it is an important task to check the on-shell consistency of the supersymmetry algebra in the sixty dimensional exceptional geometry. Note however that the complete dynamics of exceptional geometry will have to be established in order to be able to decide this question.\\ 

If the agreement of $D=11$ supergravity with a truncation of exceptional geometry is complete, then the rich symmetry structure of the former requires an explanation. The first example would be the hidden $E_{8(8)}$ symmetry of the truncated $D=11$ supergravity, which could either be related to the conformal realization of $E_{8(8)}$ \cite{GKN01b} on the $3+57$ dimensional exceptional geometry or to a generalized coset dynamics in $3+248$ dimensions. The latter would suggest an immediate extension to West's $\textit{l}_1$-representation \cite{W03}. Hence, this construction will probably provide further insights in the dynamics of the $E_{10(10)}$- and $E_{11(11)}$-conjectures \cite{DHN02,W01}. These additional structures may also fix the other numerical factors in the Lagrangian of exceptional geometry in section \ref{SummFerm}, which have been chosen so far in order to match the dynamics of $D=11$ supergravity.\\

Finally, exceptional geometry confirms the well known statement that the link between diffeomorphism-, exceptional- and supersymmetry is very tight. Therefore, exceptional geometry may possibly serve as a selection criterion for (hypothetical) supersymmetric higher curvature extensions of $d=4$ $\cN=8$ supergravity.

\mbox{}\\

{\bf Acknowledgements}\\
I would like to thank Hermann Nicolai, Axel Kleinschmidt and Bernard de Wit for refereeing my PhD-thesis \cite{Hi08} and for inspiring discussions. Furthermore, I am grateful to Thibault Damour, Marc Henneaux and Peter West for clarifying comments and valuable advice.\\

During my work, I benefitted from the support of the Studienstiftung des deutschen Volkes, of the Albert-Einstein-Institut, Potsdam, of the International Solvay Institute, Brussels and of the Universit\'e Libre de Bruxelles, where part of this work was completed.

\end{section}

\end{document}